\documentclass[aps,pra,superscriptaddress,twocolumn,amssymb,amsmath,nobalancelastpage,10pt]{revtex4-1}

\usepackage[utf8]{inputenc}
\usepackage{graphicx}
\usepackage{dcolumn}
\usepackage{bm}
\usepackage{outlines}
\usepackage{braket}
\usepackage{float}
\usepackage{makecell}
\usepackage[colorlinks,citecolor=blue]{hyperref}
\usepackage[dvipsnames]{xcolor} 
\usepackage{enumerate}
\usepackage{mathtools}
\usepackage[colorlinks,citecolor=blue]{hyperref}
\usepackage[dvipsnames]{xcolor}

\PassOptionsToPackage{hyphens}{url}\usepackage{hyperref}

\usepackage{comment} 

\makeatletter
\renewcommand{\bibsection}{%
  \par
  \baselineskip26\p@
  \bib@device{\linewidth}{82\p@}%
  \nobreak\@nobreaktrue
  \addvspace{19\p@}%
  \par
}
\makeatother

\begin{document}

\raggedbottom 
\title{A Modular Cryogenic Link for Microwave Quantum Communication\\Over Distances of Tens of Meters}

\author{Josua~D.~Schär}
\email{josua.schaer@phys.ethz.ch}
    \affiliation{Department of Physics, ETH Zurich, CH-8093 Zurich, Switzerland}
    \affiliation{Quantum Center, ETH Zurich, 8093 Zurich, Switzerland}

\author{Simon~Storz}
    \affiliation{Department of Physics, ETH Zurich, CH-8093 Zurich, Switzerland}
    \affiliation{Quantum Center, ETH Zurich, 8093 Zurich, Switzerland}

\author{Paul~Magnard}
    \affiliation{Department of Physics, ETH Zurich, CH-8093 Zurich, Switzerland}
 
\author{Philipp~Kurpiers}
    \affiliation{Department of Physics, ETH Zurich, CH-8093 Zurich, Switzerland}
    
\author{Janis~Lütolf}
    \affiliation{Department of Physics, ETH Zurich, CH-8093 Zurich, Switzerland}

\author{Melvin~Gehrig}
    \affiliation{Department of Physics, ETH Zurich, CH-8093 Zurich, Switzerland}

\author{Jean-Claude~Besse}
    \affiliation{Department of Physics, ETH Zurich, CH-8093 Zurich, Switzerland}
    \affiliation{Quantum Center, ETH Zurich, 8093 Zurich, Switzerland}

\author{Anatoly~Kulikov}
    \affiliation{Department of Physics, ETH Zurich, CH-8093 Zurich, Switzerland}
    \affiliation{Quantum Center, ETH Zurich, 8093 Zurich, Switzerland}

\author{Andreas~Wallraff}
\email{andreas.wallraff@phys.ethz.ch}
    \affiliation{Department of Physics, ETH Zurich, CH-8093 Zurich, Switzerland}
    \affiliation{Quantum Center, ETH Zurich, 8093 Zurich, Switzerland}

 \date{\today}

\begin{abstract}

Quantum technologies promise a radically new way to solve classically intractable computing problems. Superconducting circuits as a platform are at the forefront of this field.  
The cryogenic operation temperatures of superconducting circuits however impose challenges for the further scaling to many connected quantum information processing units into a local area or global network. In this work, we present a hardware solution for connecting quantum devices operating at microwave frequencies into local area networks, which enable the exchange of quantum information between spatially separated parties. Specifically, we demonstrate a modular system spanning distances of 5, 10 and 30 meters operated at cryogenic temperatures and connecting two superconducting circuit systems, located in individual dilution refrigerators, through a quantum communication channel. We develop a thermal model to evaluate the heat transfer processes in the setup, optimize the design and select appropriate materials for its construction. The assembled 30-meter-long system achieves operating temperatures of below 50 mK after a cooldown time of about six and a half days. This link enables the execution of distributed quantum computing and communication algorithms. 
It also adds the resource of non-locality, certified by a loophole\nobreakdash-free Bell test, to the field of quantum science and technology with superconducting circuits.

\end{abstract}

\maketitle

\section{INTRODUCTION}

Interconnecting multiple quantum processors to form a network is a promising approach to scaling up quantum computers~\cite{Kimble2008,Stephanie2018}. 
Quantum computing systems operating in the microwave regime, such as superconducting circuits or charges and spins in semiconductor quantum dots, require temperatures in the millikelvin regime for robustness against thermal noise and initialization in the ground state~\cite{DiVincenzo2000, Krantz2019}. Despite the recent progress in the development of microwave-to-optical transducers~\cite{Kumar2023, Brubaker2022, Stockill2022, Forsch2019, Hease2020, Mirhosseini2020, Tu2022},  
which would allow coupling microwave-based quantum processors through room-temperature optical fibers, no such system can currently combine the required conversion efficiency, bandwidth, added noise and heat loads at practical levels~\cite{Wang2022y}. At microwave frequencies, however, superconducting circuits can be interfaced with photons with near unit fidelity~\cite{Wallraff2004,Wenner2014,Pechal2014,Pfaff2017,Ilves2020,Besse2020a}. This scheme benefits from high\nobreakdash-fidelity, deterministic~\cite{Kurpiers2019,Axline2018,Zhong2021}, bidirectional~\cite{Leung2019,Zhong2019} or error correctable~\cite{Burkhart2021} quantum communication protocols. Unlike optical photons in fiber optics, microwave photons must propagate in a cryogenic environment to reduce photon loss~\cite{Kurpiers2017} and the thermal background to acceptable levels~\cite{Xiang2017, Schuetz2019}. 

In this article, we present a \textit{cryogenic link} spanning a distance of 30 meters and connecting two dilution cryostats with quantum chips inside. This unique setup enables the exchange of quantum information and the distribution of entanglement between superconducting qubits over meter\nobreakdash-scale linear distances~\cite{Magnard2020} based on single microwave photons. It also provides access to the resource of non\nobreakdash-locality available to superconducting\nobreakdash-circuit\nobreakdash-based systems~\cite{Storz2023,Storz2024,Kulikov2024}.

Building a large-scale cryogenic system spanning tens of meters comes with the challenge of optimizing the thermal performance of the setup. We discuss the choice of materials, the design of components and important aspects in the thermal engineering of the system. The insights we present are generally applicable in the development of cryogenic systems with similar requirements~\cite{Alduino2019,Hollister2017}.

The article is structured as follows. In Section~\ref{sec:modular_design_concept} we present the basic concept and general structure of the modular cryogenic quantum link, and outline the functions of individual modules. We review the most important design criteria in Section~\ref{sec:design_optimization}, describing the specific design choices we made to realize our cryogenic link. We discuss the performance of cooldowns of systems with a length of 5, 10 and 30 meters in Section~\ref{sec:cooldown_results}. We further estimate distance scales potentially achievable with this type of setup. Finally, we summarize and discuss potential system improvements in Section~\ref{sec:summary}.

\section{MODULAR DESIGN CONCEPT}
\label{sec:modular_design_concept}

\begin{figure*}
	\begin{center}
		\includegraphics[width=2.0\columnwidth]{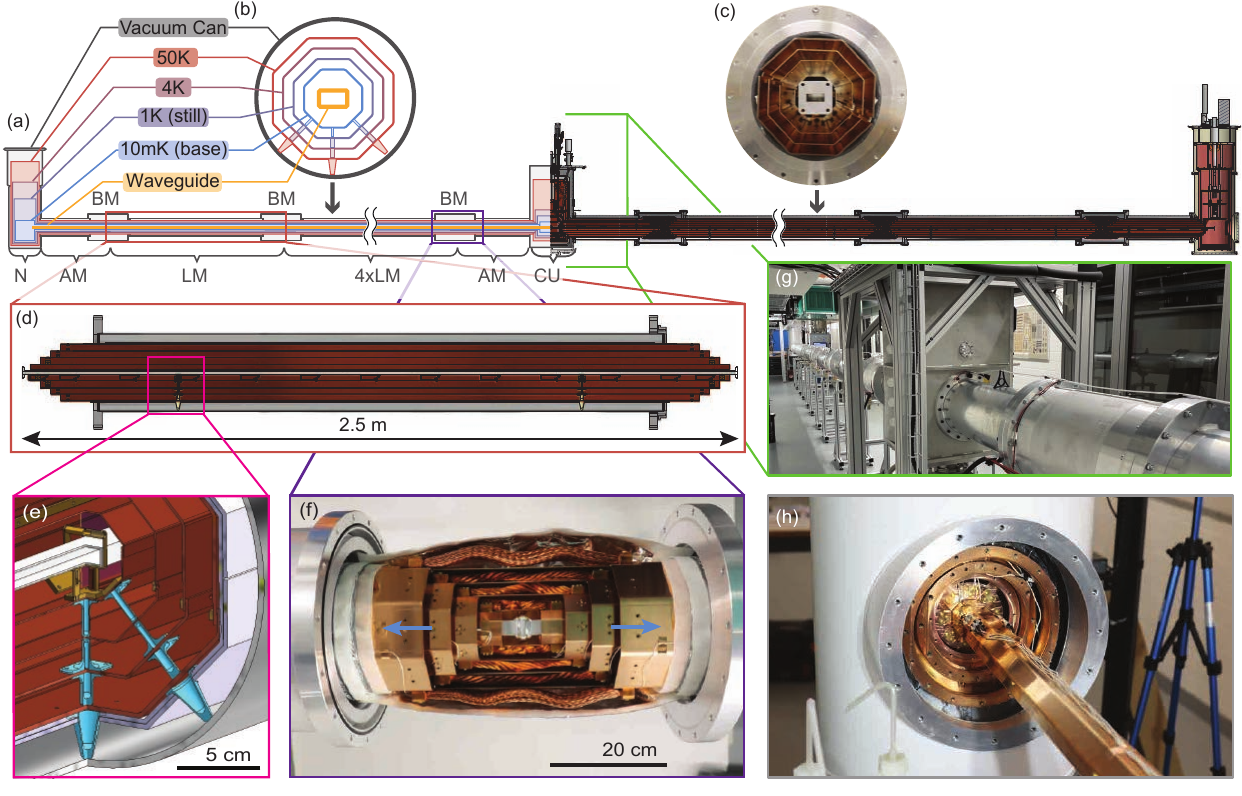}
		\caption{
			(a) Longitudinal cross-section of a schematic representation (left half) and a 3D model (right half) of a $30\textrm{\,m}$ cryogenic link system. The link is composed of four adapter modules (AM), 12 braid modules (BM), 10 link modules (LM), one cooling unit (CU), and two nodes (N). (b) Schematic representation, and (c) photograph of a transverse cross-section of a link module. (d) Longitudinal cross-section of the 3D model of a link module. (e) Enlarged cross-sectional view of a set of posts (blue), which provide mechanical support for the radiation shields. (f) A top view photograph of a braid module connecting two link modules. For demonstration purposes, only the lower half of the braid module is assembled and closed with radiation shields, braids and multi-layer insulation. The blue arrows indicate the direction in which the copper shields of the link modules move during a cooldown due to thermal contraction, see discussion in Section~\ref{thermal_expansion_subsec}. (g) A photograph of a part of the $30$-m-long cryogenic setup close to the cooling unit in the middle. (h) A photograph showing the flanges of a node with an attached base shield of an adapter module. 
		}
		\label{fig:design_concept}
	\end{center}
\end{figure*}

To realize quantum communication in a cryogenic microwave channel, we engineer, assemble and demonstrate a distributed modular cryogenic system capable of housing superconducting qubits at each node. The system is composed of two dilution refrigerators connected through a cryogenic link with an additional pulse tube cooler positioned in the middle of the system, see Fig.~\ref{fig:design_concept}(a). We constructed the cryogenic link from several individual modules to facilitate the manufacturing, assembly, maintenance, and extension of the system.

The design of the modules of the cryogenic microwave link is inspired by that of a standard dilution refrigerator.
A dilution refrigerator~\cite{Pobell2006} typically contains four temperature stages. The two higher temperature stages are cooled to approximately $50\textrm{\,K}$ and $4\textrm{\,K}$ with a two-stage pulse tube cooler [see Fig.~\ref{fig:design_concept}(a)]. Millikelvin temperatures are reached with a $^3\textrm{He}/^4\textrm{He}$ dilution unit, effectively cooling the still stage to about $1\textrm{\,K}$ and the base temperature stage to approximately $10\textrm{\,mK}$. Radiation shields screen thermal radiation from higher temperature stages and are made out of copper or aluminium. To provide thermal isolation of the stages from the room-temperature environment, dilution refrigerators are typically housed in an O\nobreakdash-ring\nobreakdash-sealed aluminium vacuum container which is initially evacuated to $ 10^{-4} \textrm{\,mbar} $ to suppress heat exchange by convection of residual gas.

As in a dilution refrigerator, each cryogenic module of our microwave quantum link consists of a set of concentric copper radiation shields enclosed by a vacuum can at room temperature, Figs.~\ref{fig:design_concept}(b),(c). In this arrangement we achieve the aforementioned typical dilution\nobreakdash-refrigerator temperatures also along the link.

For quantum communication between the nodes we use a rectangular aluminium WR90 waveguide~\cite{WR90} (specified in the Electronic Industries Alliance standards). An alternative approach requiring less space is to route superconducting coaxial cables~\cite{Renger2023} between the nodes. However, in general, such cables are more lossy than waveguides of the same material~\cite{Kurpiers2017}. Our waveguide is housed inside the innermost radiation shield cooled to the lowest temperature achievable in our system. We mount a superconducting circuit in its dedicated sample package at each node to perform quantum experiments~\cite{Magnard2020,Storz2023,Storz2024,Kulikov2024}. We connect the ends of the rectangular waveguide to the superconducting circuits using a waveguide to coaxial adapter and a coaxial microwave cable.

We designed and manufactured four principal modules to construct the entire link assembly connecting the two nodes of the system.

\textit{Link modules} are the primary building block of the cryogenic link, covering most of the distance between the nodes. A module spans a distance of $2.5\textrm{\,m}$ when measured from the waveguide flange on one side to the corresponding one on the other end. 
Figure~\ref{fig:design_concept}(d) shows a cross\nobreakdash-sectional CAD drawing of a link module. The copper radiation shields are shorter than the waveguide on both sides, with their lengths staggered in decreasing order from base temperature radiation shield to the aluminium vacuum can. This design ensures good accessibility of each stage during the assembly of the link. 

The radiation shields within a link module are separated from and mechanically supported by each other using two sets of three thin\nobreakdash-walled posts [Fig.~\ref{fig:design_concept}(e)]. 
The waveguide is held in place inside the base temperature shield by brass holders, see yellow parts around the waveguide in Figure~\ref{fig:design_concept}(e). 

We designed the radiation shields with an octagonal cross\nobreakdash-section rather than a cylindrical one to facilitate the mounting of elements on the shields, such as the flanges of the thermal braids, mechanical supports, or temperature sensors. The octagonal shields are assembled from two half\nobreakdash-shells joined with screws along the link axis. Using welded shields instead would reduce leakage of thermal radiation, but reduce the accessibility to mount components on the shield like support posts.

An \textit{adapter module} follows the same concept as a link module, where one end comes with the same staggering profile of the radiation shields as in a link module, while the other end is adapted to connect to the cryostat flanges. These flanges are a custom modification of a standard \textit{Bluefors} LD dilution refrigerator [see Fig.~\ref{fig:design_concept}(h)].

\textit{Braid modules} are used to connect link modules to each other or adapter to link modules. Its function is to allow for movements of mechanical components with respect to each other, which are induced by thermal contraction of radiation shields, while providing good thermal conduction between the modules. This is achieved by a set of 5 to 8 flexible and highly conductive copper braids on every temperature stage. 
The gap between any two adjacent radiation shields is in addition closed by short copper shields installed below the copper braids to screen thermal radiation [see Fig.~\ref{fig:design_concept}(f)]. To assemble a braid module, we close each temperature stage separately starting with the base temperature and ending with the vacuum can. The staggered lengths of the link and adapter modules provide the space required for the attachment of the braids.

We position a \textit{cooling unit}, containing a pulse tube cooler without a dilution unit, between the nodes of the cryogenic link to bridge distances $\gtrsim 20\textrm{\,m}$ [see Fig.~\ref{fig:design_concept}(a),(g)]. The cooling unit provides additional cooling power required to operate the system, without the cooling unit the $50\textrm{\,K}$ radiation shields in the middle of the link would become too hot and heat up the colder temperature stages considerably, causing the cooling to fail; see Section~\ref{sec:cooldown_results}. A description of the custom cooling unit that we have designed and built is in Appendix~\ref{intermediate_cooler}.

To assemble the cryogenic link, we first position the nodes and cooling units at the desired location in the laboratory and attach the corresponding adapter modules. Then we position and align each link module mounted on stable supports one by one between the adapter modules. Finally, we close the connections between the segments by assembling the braid modules. Methods to ensure good thermal contact and discover radiation leaks are discussed in Appendix~\ref{production_assembly}.

\section{DESIGN AND MATERIAL OPTIMIZATIONS}
\label{sec:design_optimization}

A key aspect in the successful engineering of a cryogenic link is the understanding of heat transfer in the system, which generally consists of two counteracting contributions. First, every temperature stage is exposed to \textit{heat load} stemming from thermal radiation between radiation shields and from finite thermal conduction through the mechanical support structures. Both effects deliver heat from higher to lower temperature stages of the system.
Second, this distributed heat load generates a \textit{heat flow} along the radiations shields transporting thermal energy towards heat sinks located at the nodes and the cooling unit. This heat flow generates a temperature gradient inside the shields and braids, and temperature drops at contact interfaces e.g.~between the modules of the link.
In steady state, the heat load is in balance with the heat flow towards the sink at every position and temperature stage along the link, defining the temperatures of individual elements.

An additional constraint in designing the cryogenic system is the material\nobreakdash-dependent thermal contraction of its components during each cooldown and warm\nobreakdash-up, which requires a mechanically flexible structure between the modules and different temperature stages to avoid physical damage.

\subsubsection{Decreasing the heat loads}

We quantitatively want to estimate the radiative heat load impinging on each temperature stage of the system and compare it to the cooling power available at the nodes. We calculate the thermal radiation according to the Stefan-Boltzmann law, see Eq.~\eqref{eq:radiation} in Appendix~\ref{thermal_model} for typical temperatures i.e., from base shield ($0.01\textrm{\,K}$) over still ($1\textrm{\,K}$) and 4K shields ($4\textrm{\,K}$) to the 50K shield ($50\textrm{\,K}$) with the laboratory including the vacuum can at ($293\textrm{\,K}$). For the Copper shields and the Aluminium can we use typical emissivities of $\epsilon_{\textrm{Cu}} \simeq 0.025$ and  $\epsilon_{\textrm{Al}} \simeq 0.05$~\cite{Lienhard2020, Bramson1968}. The radiative heat load on each temperature stage is on the order of $(10^{-9},10^{-6},0.01,10)\textrm{\,W/m}^2 $ from base to 50K. At the temperatures mentioned above, the nodes have a cooling power of $(4*10^{-6},2*10^{-2},0.5,14)\textrm{\,W}$, [see Table~\ref{tab:all_parameters} in Appendix~\ref{thermal_parameters}].

Below the 50K stage the cooling power is orders of magnitude larger than the radiative heat load per unit area. But on the 50K stage for a 30\nobreakdash-meter\nobreakdash-long cryogenic link  with a surface area of ${\sim20\textrm{\,m}^2}$ the radiative heat load is around $200\textrm{\,W}$. Using only copper shields would heat up the 50K stages in the nodes to over ${\sim250\textrm{\,K}}$. Each node has an internal coldtrap installed on the 50K stage, which is used to capture and remove contaminants like gases and water vapour in the dilution unit. For proper operation, these coldtraps must remain below $77\textrm{\,K}$, which is not possible with such a high radiation heat load.

In order to overcome this large heat load, we install multi\nobreakdash-layer insulation (MLI)~\cite{Parma2015,marlow_design_2011} on the outer surface of the 50K radiation shield. With this measure, we decrease the emissivity of $\varphi_{\textrm{rad}}\approx 6.4\textrm{\,W/m}^2$ to $\varphi_{\textrm{rad}} \approx 1\textrm{\,W/m}^2$ (see Appendix~\ref{thermal_properties_link}). The reduced heat load enables efficient cooling of the 50K temperature stage below $77\textrm{\,K}$ at the nodes. The result shows that the use of MLI is a crucial element to lower the radiation heat loads on the 50K radiation shield. 

\begin{figure}[t]
	\begin{center}
		\includegraphics[width=1.0\columnwidth]{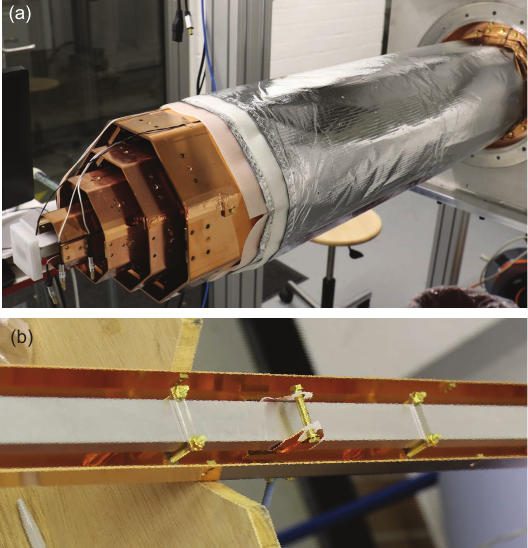}
		\caption{(a) Photograph of an adapter module connected to the cooling unit of the $30\textrm{\,m}$ link with the multi layer insulation installed on the 50K shield (silver foil). (b) Detailed photograph of the waveguide installed inside the open base temperature shield.  
		}
		\label{fig:optimizations}
	\end{center}
\end{figure}

The MLI itself consists of a set of plastic foils coated with a thin aluminium layer as reflective surfaces separated from each other by thin polyester or glass fiber nets. The principle of these multiple layers is to reflect back a large percentage of the irradiated heat load at each layer~\cite{MLI, Kang1999}. Figure~\ref{fig:optimizations}(a) shows the installed customized MLI with 30 layers in total~\cite{Shu1998}. The two blankets of MLI with 15 layers each are wrapped around the shield, and fixed with hook\nobreakdash-and\nobreakdash-loop fasteners. The fasteners of the two blankets are designed not to overlap, reducing radiation leakage through them, see white stripes at the ends of the MLI in Figure~\ref{fig:optimizations}(a). We covered the remaining gap between two adjacent MLI blankets with reflective aluminium tape to reduce heat leaking through the gaps~\cite{Johnson2009,Finckenor1999}.

The second contribution to the heat load originates from conduction through the support posts [see Fig.~\ref{fig:design_concept}(e)].
We minimize this conductive heat load by selecting \textit{Bluestone}, a 3D printed nano\nobreakdash-composite material in a stereo lithography process followed by a tempering step~\cite{accura_bluestone}.
Compared to other tested materials, Bluestone has the highest yield strength to thermal conductivity ratio among the materials we investigated (see Appendix~\ref{thermal_properties_bluestone}).

 \subsubsection{Optimizing heat transport}

In order to minimize the temperature gradient along the link and thus minimize the temperature in the middle between the cooling nodes, we construct the radiation shields using high\nobreakdash-purity copper, which is one of the best cost\nobreakdash-effective thermal conductors at cryogenic temperatures~\cite{Pobell2006}. 
In addition, surface oxides can be removed from copper surfaces efficiently before installation~\cite{Woodcraft2005a}. The surfaces are stable in dry and inert atmospheres as in the vacuum can, maintaining low emissivity and low contact resistances when mated with other copper parts.

Three common types of copper have high thermal conductivity, ranging in purity from $99.9\%$ to $99.99\%$: Electrolytic-tough-pitch (ETP) copper, oxygen-free (OF) copper and oxygen\nobreakdash-free electronic (OFE) copper. The impurity limits for other elements are specified differently in the corresponding standards~\footnote{German copper association: \href{https://standalone.kupferschluessel.de/standalone.php?lang=english}{Kupferschlüssel 1.0} (visited on 06.06.2024)}.
 
We measured the thermal conductivity $\rho$ of these three pure copper grades to select a suitable type. At room temperature, $\rho$ differs only by a few per mil for these types, but at cryogenic temperatures the level and type of impurities in the material becomes relevant and $\rho$ can vary by orders of magnitude~\cite{Pobell2006,Fickett1982} (see Appendix~\ref{RRR_copper}).
To extract the thermal conductivity $\rho$, we placed copper test strips in a vacuum chamber and cooled them using liquid helium to approximately 4.2~K, the method is described in detail in the Appendix~\ref{RRR_copper}.

In summary, the OFE copper available to us showed a conductivity of $\rho_{\textrm{OFE}}\approx 1400 \textrm{\,W/(Km)}$ at $4.2\textrm{\,K}$, significantly higher than ETP $\rho_{\textrm{ETP}}\approx 700 \textrm{\,W/(Km)}$ or the best OF copper $\rho_{\textrm{OF}}\approx 500 \textrm{\,W/(Km)}$ we measured (see Appendix~\ref{RRR_copper} for details).

In addition to material selection, the thicknesses of the four radiation shields\textemdash and thus the thermal conductance\textemdash are adjusted to the expected heat load. The largest heat load is on the 50K shield, mainly due to the thermal radiation, and the smallest for the base temperature shield. Heavier shields need thicker support posts, which in turn leads to higher heat load from the posts. We chose thicknesses of $3\textrm{\,mm}$ for 50K, of $2\textrm{\,mm}$ for 4K, and of $1\textrm{\,mm}$ for the two lowest temperature shields~\cite{Kurpiers2019a}.

Each mechanical interface between a radiation shield and a braid creates a thermal contact resistance. In a test setup (see Appendix~\ref{thermal_properties_link}) we observed that at cryogenic temperatures the contact resistance depends on the temperature as $ R_{\textrm{contact}} \propto T^{-2}$~\cite{Salerno1997}, while at the same time below $\sim 10\textrm{\,K}$ the characteristic thermal resistance of braids (bulk copper) scales as $R_{\textrm{bulk}} \propto T^{-1}$~\cite{Pobell2006}. As a consequence, the contact resistance contributes significantly at the 4K stage and dominates over the bulk resistance of the braids at still and base temperatures. Moreover, the contact resistance also depends on the force pressing the two parts together, on the thickness of the oxide layer and on the surface roughness~\cite{Salerno1997}. Therefore, contact resistances are best determined in test measurements close to the real conditions by using the same components, manufacturing processes and assembly.

In general, to efficiently thermalize parts or devices below 50K stage, we minimize the number of contact interfaces along the cooling path. Our modular link introduces two contact interfaces on every temperature stage for each braid module, and one for each adapter module. We utilise the available space of the braids to fit as many screws as possible and tighten the screws with  appropriate torque. We have electropolished the contact surfaces and added \textit{Apiezon N} grease with the intention to reduce thermal contact resistances at these interfaces, which can significantly lower $R_{\textrm{contact}}$ as discussed in~\cite{Salerno1997}.

 \subsubsection{Thermal expansion}
 \label{thermal_expansion_subsec}

Copper and aluminium parts contract up to $ 3.25 \textrm{\,mm/m} $~\cite{Pobell2006} and $ 4.15 \textrm{\,mm/m} $~\cite{nist_Al} between room and millikelvin temperatures. As a consequence, the length of a base temperature radiation shield in a link module shrinks by about $8 \textrm{\,mm}$. Notably, the cooling rates and final temperatures of the temperature stages differ, leading to different thermal contraction.

For the design of the link, it is therefore important to consider this effect to avoid mechanical stress and potential damage in the support structure. Accordingly, we fixed the support posts in the link module to only one of the adjacent shields, allowing the two shields to slide against each other in the longitudinal direction of the link and thus eliminating thermal stress in the posts [see Fig.~\ref{fig:design_concept} (d),(e)]. As described in Sec. \ref{sec:modular_design_concept}, braid modules provide flexible, high  thermally\nobreakdash-conductive connections between any two link modules. These braids are designed with an excess length of $12\textrm{\,mm}$ at room temperature to allow for the thermal contraction of the radiation shields, indicated with the blue arrows in Figure~\ref{fig:design_concept}(f), without creating mechanical stress. The short copper shields underneath the braids are attached with screws to a radiation shield in one link module for thermalization, and the other end can freely slide over the radiation shield of the other link module.

Thermal contraction further imposes a challenge for the aluminium waveguide, consisting of individual sections, connected rigidly with screws to form a low\nobreakdash-loss quantum channel. 
With a total contraction of $125\textrm{\,mm}$ along the distance of 30 meters during a cooldown, we ensure a sufficient mobility of the waveguide along the longitudinal direction in the base temperature shield. For this purpose we use a brass stand covered with a Teflon sheet [see Fig.~\ref{fig:optimizations} (b)], which, thanks to its low friction, prevents the waveguide from sticking and exerting stress on the base temperature radiation shields. 
We use copper braids with an excess length of $150 \textrm{\,mm}$ to thermalize the waveguide and allowing it to slide inside the base temperature shield. This leaves a margin of $25\textrm{\,mm}$ compared to the expected contraction of the waveguide in the 30\nobreakdash-meter\nobreakdash-long link. 
Finally, we use flexible coaxial microwave cables to connect the waveguide to the superconducting circuits installed in the nodes (see Section \ref{properties_quantum_channel}).

Aluminium becomes superconducting below $T_c=1.2 \textrm{\,K}$~\cite{Satterthwaite1961}. With this phase transition, the thermal conductivity drops sharply and can reach values of $\rho_{\textrm{AL}}=10^{-6}\textrm{\,W/(Km)}$ at $50\textrm{\,mK}$ comparable to thermal insulators~\cite{Pobell2006}. We therefore thermalise the waveguide to the base temperature shield with copper braids every $0.20 \textrm{\,m}$ [see Fig.~\ref{fig:optimizations} (b)].

\section{THERMAL PERFORMANCE OF THE LINK}
\label{sec:cooldown_results}

Over the course of the development of the cryogenic link, we have assembled and cooled down three systems bridging distances of $5\textrm{\,m}$, $10\textrm{\,m}$ and $30\textrm{\,m}$. The cooldown of a $5\textrm{\,m}$ and a $10\textrm{\,m}$ link served as an initial assessment of the modular system performance~\cite{Magnard2020} and its scalability to the length of $30\textrm{\,m}$ targeted to enable a loophole\nobreakdash-free Bell test with superconducting
circuits~\cite{Storz2023}.
The steady\nobreakdash-state temperature profiles recorded during cooldowns of the 5 and 10-m systems allowed us to model the thermal processes in the link and plan for the 30\nobreakdash-meter\nobreakdash-long system, see Appendix~\ref{thermal_model}.

\subsubsection{Measured temperature profiles}

\begin{figure*}
	\begin{center}
			\includegraphics[width=2.0\columnwidth]{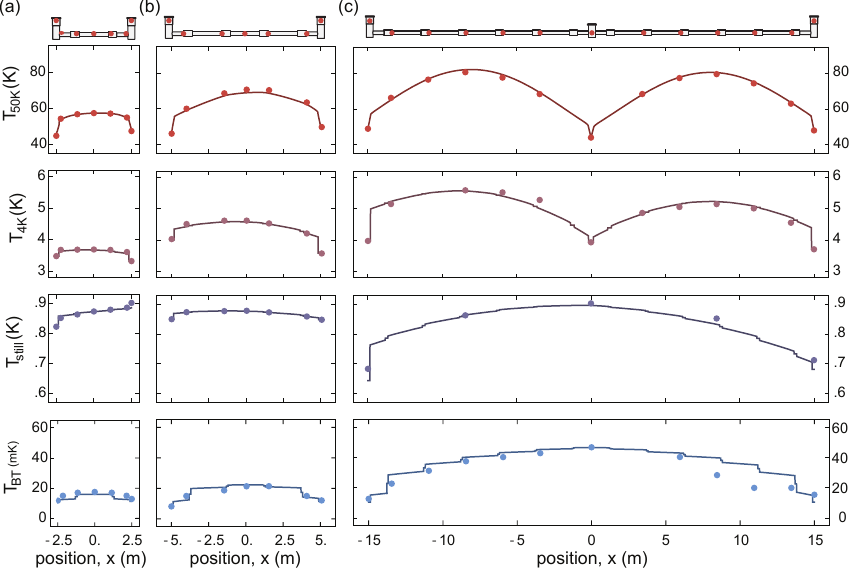}

		\caption{
			Steady-state temperature distribution of the (a) 5-m-long, (b) 10-m-long and (c) 30-m-long cryogenic link. The pictograms above the plots represent the cryogenic prototypes and the temperature sensor positions (dots) in scale with the x-axis. The  solid lines are results of the thermal model simulations.
		}
		\label{fig:temperatures}
	\end{center}
\end{figure*}

To monitor the temperature profile during each cooldown, we installed up to 48 temperature sensors along the cryogenic link (see Appendix~\ref{thermometry}). Figure~\ref{fig:temperatures} (a-c) shows the measured steady\nobreakdash-state temperatures on all temperature stages along the cryogenic link for the 5, 10, and 30-m systems. Due to the thermal diffusivity of copper, a property characterizing how fast a material can transport heat away~\cite{Lienhard2020}, the 50K stage takes the longest to reach steady state. The temperatures were recorded 4 to 7 days after condensing the $^3\textrm{He}/^4\textrm{He}$ mixture and reaching millikelvin temperatures on the base plates in the nodes.
As expected, the temperature profiles in general exhibit a convex shape with minima at the cooling points. We find that on all stages the temperatures and the temperature gradients along the link increase with the length of the system. Over the entire 30-meter-long system, the 50K stage remains below $80\textrm{\,K}$, reaching its maximum at the midpoint between the nodes and the cooling unit. A similar observation is made for the 4K stage, which stays below $6\textrm{\,K}$. The 4K and still stage temperature profiles show some asymmetry at the nodes. We attribute this difference to the thermal-shield design improvements made by the manufacturer for the dilution cryostat on the right-hand-side node, which was acquired separately and later than the left-hand-side one.

The base temperature along the link remains below $50\textrm{\,mK}$ everywhere and for all three systems. At the nodes base temperatures of $10$ to $15\textrm{\,mK}$ are reached, which is similar to the typical values in a single dilution refrigerator.

The largest temperature gradients between two subsequent modules are found between the heat sink in the nodes and the adapter module from the 50K to the still stage. This is due to a less optimized selection of materials for the dilution refrigerator radiation shields, which show a lower thermal conductivity than the copper used in the link. The interfaces between the radiation shields of the dilution refrigerator and the mounted flanges connecting the adapter module to the node cause significant contact resistance~\cite{Magnard2021}. A possible reason could be the stainless steel screws used to assemble these parts, because the thermal contraction of stainless steel is lower than the one of copper, causing to lower the preload when cooling down.

In addition to measuring the temperatures directly, we simulate the temperature profiles and compare the results. In a first step, we use a finite element software and a 3D model of the system to get the steady\nobreakdash-state temperature distribution of the 50K stage. We use the independently determined material properties listed in Appendix~\ref{thermal_parameters} Table~\ref{tab:all_parameters} as input parameters. We set the vacuum can temperature to $293\textrm{\,K}$ and the temperatures at the nodes of the 50K stage to their measured values as the only boundary conditions. 
The finite element method calculates the temperature distribution in the 3D model iteratively to approximate the temperature profile in the 50K stage; it also takes the temperature of the stage itself into account to determine the radiative and conductive heat load.

In a second step, using the obtained temperature profile of the 50K stage, we recursively simulate the temperature profiles of the lower temperature stages. We use numerical calculations and the thermal model in Appendix~\ref{thermal_model} with the material properties as input parameters and the temperature profile of the next hotter stage as boundary condition. 

The simulated temperature profiles are shown with solid lines in Figure~\ref{fig:temperatures}. Overall, our model is in good agreement with the temperatures measured in all three systems. We observe discrete jumps in the modeled temperatures at the interfaces between modules due to thermal contact resistance, relevant at the still and base stages. There is a minor discrepancy between the measured and simulated temperatures for the base stage on the right-hand-side of the 30-m-long link. This difference could arise from calibration errors in the temperature sensors, which we calibrated ourselves, or from other factors such as variations in the contact resistance.

\subsubsection{Cool down and dwell times}

The  time to cool down to the base temperature of slightly below $20\textrm{\,mK}$ at the node cryostats of links with lengths 5, 10 and 30 meters is 2, 3.5 and 6.5 days, respectively. For comparison, a single dilution cryostat needs around 1.5 days. With each of the three different\nobreakdash-length systems, the condensation process of the $^3\textrm{He}/^4\textrm{He}$ mixture, necessary to reach millikelvin temperatures, from pulse tube temperatures ($\approx 4\textrm{\,K}$) took less than eight hours, similar to a dilution cryostat. The cool down dynamics of the 30 meter system is discussed in more detail in Appendix~\ref{cooldown_dynamics}.

After about two weeks, the  30\nobreakdash-meter\nobreakdash-long system reaches the lowest temperature configuration. Beyond that, the temperature of the 50K shield begins to slowly increase with time at a rate of about $ 0.2 \textrm{\,K/day}$ in the center between a node and the cooling unit, and $ 0.1 \textrm{\,K/day}$ at the nodes. We assume the increase in the temperature of the 50K stage is due to the gas diffusion through the O\nobreakdash-rings of the vacuum can~\cite{McKeen2012}, and we hypothesize that this gas condenses on the surface of the MLI on the 50K shielding, causing its reflectivity to decrease over time~\cite{Renger2023}.
This effect does limit the operation time of the system by causing too much heat load on the 50K stage or by substantially heating up the underlying 4K stage by increased thermal radiation from the hot 50K shield and heat load through the support posts. With O\nobreakdash-ring sealed vacuum flanges the hold time of the 30\nobreakdash-meter\nobreakdash-link is about 6 months. This time could be increased by using better seals, such as the ones used in the CF flanges standard for ultra high vacuum ($<1*10^{-7}\textrm{\,mbar}$)~\cite{Pfeiffer}. We also observe that warming up the system to $\approx230 \textrm{\,K}$ and evacuating the vacuum can restores the system and allows for the following cooldown to proceed nominally.

\subsubsection{Properties of the quantum channel}
\label{properties_quantum_channel}

For quantum communication experiments~\cite{Magnard2020,Storz2023,Storz2024,Kulikov2024}, we use the transversal electric mode $\textrm{TE}_{10}$~\cite{Pozar2021} of a WR90 rectangular waveguide made from aluminium as a microwave quantum channel.
The attenuation factor of the waveguide was measured to be $\alpha < 1\textrm{\,dB/km}$~\cite{Storz2023a}, using the same methods as in~\cite{Magnard2021,Kurpiers2017}. The loss in the 30-meter-long channel is therefore lower than $0.03\textrm{\,dB}$.
In current experiments~\cite{Storz2024,Kulikov2024}, the photon transfer loss ($0.55$ to $ 0.65\textrm{\,dB}$) is likely dominated by the connection between the superconducting chip, the sample package and the rectangular waveguide on each node. It incorporates transitions from on\nobreakdash-chip coplanar waveguides to a printed circuit board via wirebonds, then a sub\nobreakdash-miniature push\nobreakdash-on (SMP) connector is soldered on the printed circuit board to route the photon to a coaxial cable. The cable is connected to the end of the rectangular waveguide using a waveguide\nobreakdash-to\nobreakdash-coax adapter. Some of the components are not superconducting and in total they cause more loss than the waveguide itself~\cite{Storz2023a}.

\subsubsection{Temperature profile extrapolation}
Having constructed and carefully characterized cryogenic links of three different lengths, we are now in position to extrapolate the temperature profile simulations for longer cryogenic link systems. Specifically, we are interested in estimating the maximum length of a cryogenic link one should be able to realize using the architecture described in this paper. We determine this length by extracting thresholds from the developed thermal model and based on the three criteria forming a prerequisite for a successful system cooldown and operation:
\begin{enumerate}[i]
   
    \item The highest temperature of the $4 \textrm{\,K}$ plate, at which we know from internal experiments that Helium can be successfully condensed (using \textit{Bluefors} LD series dilution refrigerators), is $5.2 \textrm{\,K}$. This temperature must therefore not be exceeded at the end nodes.
     \item The still plate in a dilution refrigerator must remain below $1.2 \textrm{\,K}$. As this limit is approached, we observe the base-plate temperature rising and becoming unstable.

    \item To assert low thermal population~\cite{Jin2015b,Kulikov2020} of the qubits operated in the dilution refrigerators at the nodes, we target base plate temperatures below $50 \textrm{\,mK}$.

\end{enumerate}

While the first two criteria are necessary for the operation of our dilution refrigerators and must be met, the third may differ depending on the application, reset schemes~\cite{Magnard2018} or communication protocols used~\cite{Vermersch2017,Xiang2017}.

\begin{figure}[b]
	\begin{center}
			\includegraphics[width=1.0\columnwidth]{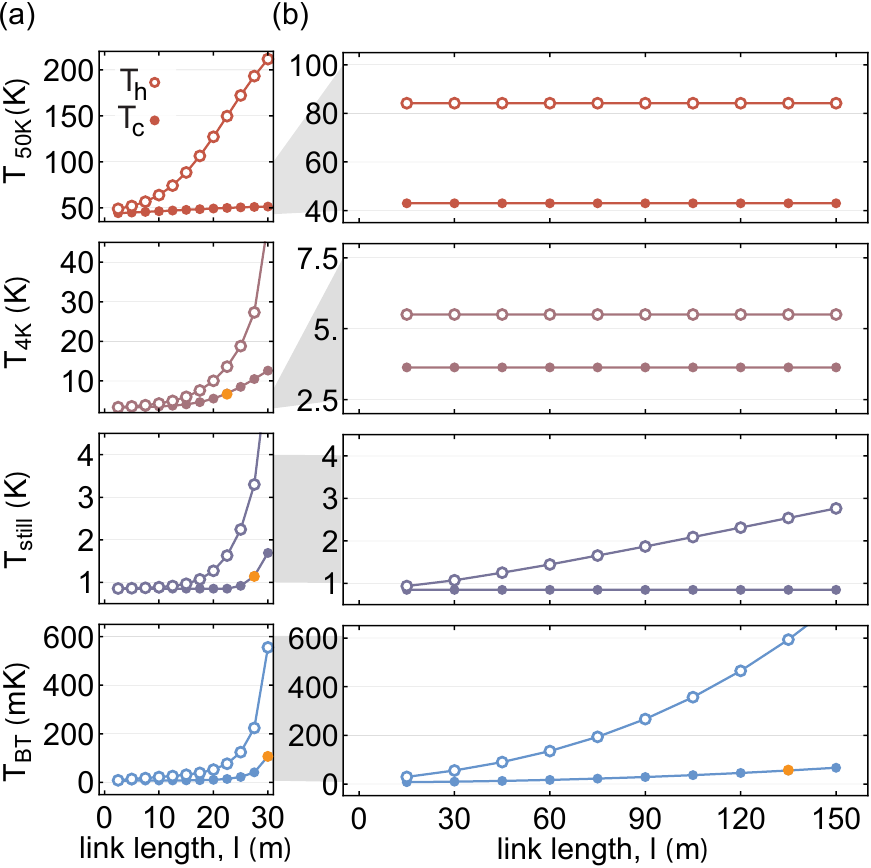}

		\caption{
			Lowest (solid dots) and highest (open dots) temperatures along the cryogenic link {\it vs} link length determined from our thermal model (see Appendix~\ref{thermal_model}) for: (a) a cryogenic link without additional cooling units, (b) a cryogenic link with additional cooling units installed every $15\textrm{\,m}$. The orange dots highlight the length at which a performance criterion for successful link operation is violated (see main text). The grey trapezoids between figures (a) and (b) visualize the temperature rescaling.
		}
		\label{fig:length_limits}
	\end{center}
\end{figure}

Given these criteria, we first determine the temperature profile of link systems at lengths from $2.5 \textrm{\,m}$ to $30 \textrm{\,m}$ without additional cooling units installed along the length of the link. The resulting temperatures at the nodes $T_c$ and the highest ones in the middle of the link $T_h$ at all temperature stages are displayed in Figure~\ref{fig:length_limits} (a). Both $T_c$ and $T_h$ increase with the link length $l$, while $T_h$ is more affected. The $4 \textrm{\,K}$ plate exceeds $5.2 \textrm{\,K}$ (criterion (i)) around a length of $20\textrm{\,m}$ marked in Fig.~\ref{fig:length_limits} (a) with an orange dot. Criteria (ii) and (iii) for still and base temperature fail for lengths beyond $20\textrm{\,m}$ due to significant temperature increase on all stages.

To bridge larger distances with our cryogenic link architecture, additional cooling units between the nodes, cooling the 50K and 4K stages, are required. Our simulations suggest that with cooling units installed every $15\textrm{\,m}$ along the cryogenic link, the system could be extended to a length of $120 \textrm{\,m}$ until criterion (iii) on the base stage is violated [see Fig.~\ref{fig:length_limits} (b)]. This would, for example, allow the realization of a cryogenic link between neighboring buildings using the presented modular design.

\section{SUMMARY AND OUTLOOK}
\label{sec:summary}

In this work, we presented a modular cryogenic link for microwave quantum communication connecting two dilution refrigerators. We discussed design aspects essential for the realization of a cryogenic link with a length of up to $30\textrm{\,m}$. The modular design facilitates the manufacturing and engineering process as well as the assembly of the system. Our 30-m-long cryogenic link reaches base temperature of less than $50\textrm{\,mK}$ along the entire length of the link. With the cooldown time not exceeding one week, it provides up to half a year of continuous operation with a stable base temperature profile.

The excellent cryogenic performance of our system demonstrates the robustness of the modular design. The good agreement between the simulation of the thermal model and the measured temperature profile shows evidence that the extracted thermal parameters are reliable and the simplified thermal model is a very good approximation for the complex heat transfers in the system.
We further used the model to estimate the feasible length of cryogenic links, showing that cryogenic networks with a length of up to $120\textrm{\,m}$ are achievable by using additional cooling units at the 50K and 4K stages every $15\textrm{\,m}$. Further extension to hundreds of metres is expected to be possible by adding extra dilution refrigerators along the length of the link. 

Building a quantum local area network connecting several nodes in a grid over tens of meters without additional cooling units could be achieved with further optimization of the design and material selection. For example, the radiation heat load on the 4K stage can be lowered by installing multi-layer insulation. To achieve larger thermal conductivity along the link, the copper radiation shields can be oxygen annealed after manufacturing~\cite{Fickett1982,Rosenblum1977}.

The cryogenic link was successfully used as a quantum communication channel with a length of $5\textrm{\,m}~$\cite{Magnard2020} and extended to $30\textrm{\,m}$ for a loophole-free Bell test~\cite{Storz2023} and two successive device-independent experiments utilizing it: complete self-testing of a system of remote superconducting qubits~\cite{Storz2024} and device-independent randomness amplification~\cite{Kulikov2024}. This suggests that our modular design can be used to realize a setup for quantum networks~\cite{Xiang2017,Penas2022}, for the exploration of device\nobreakdash-independent physics~\cite{Vazirani2014,Pironio2010,Kessler2017,Supic2019a,Bowles2018}, and for non\nobreakdash-Markovian waveguide QED~\cite{Dinc2020,Dinc2019,Calajo2019b}.

\section*{ACKNOWLEDGMENTS}

We thank Reto Schlatter, Niti Kohli, and Alain Fauquex for laboratory space preparation, pre\nobreakdash-assembly of the 30\nobreakdash-m\nobreakdash-link modules and ordering of components; Raphael Keller and Adrian Schwarzer for contributions to early prototype design and realization; Fabian Marxer, Manuel Hindering, Simon Wili and Theo Walter for assistance with prototype assembly and thermal analysis; Nicholas Meinhardt for contributions to waveguide loss measurements and thermalization design; Bahadir Dönmez, Milan Lezaic, Kirill Feldman, and Josef-Anton Agner for technical support. We thank Gianni Blatter for valuable comments on the manuscript. We also thank the workshop teams of the Department of Physics of ETH Zurich and the Paul Scherrer Institute (PSI) for the fast and reliable production of mechanical components of the cryogenic setup. This work was supported by the European Research Council (ERC) through the “Superconducting Quantum Networks” (SuperQuNet) project, by the European Union’s Horizon 2020 FET-Open project SuperQuLAN (grant no. 899354), by the National Centre of Competence in Research “Quantum Science and Technology” (NCCR QSIT), a research instrument of the Swiss National Science Foundation (SNSF), and by ETH Zurich.

\appendix

\normalsize

\section{CUSTOM BUILT RADIATION SHIELD COOLING SYSTEM}
\label{intermediate_cooler}

As explained in Sec.~\ref{sec:modular_design_concept}, operating a 30-meter-long cryogenic link at sufficiently low temperatures requires installing an additional cooling unit in the center of the system. For this purpose, we designed and assembled a pulse-tube based cryostat to provide additional cooling power at the 50K and 4K temperature stages at a desired location along the link. In this section, we discuss the main features of this cooling unit.

\begin{figure*}
	\begin{center}
			\includegraphics[width=2.0\columnwidth]{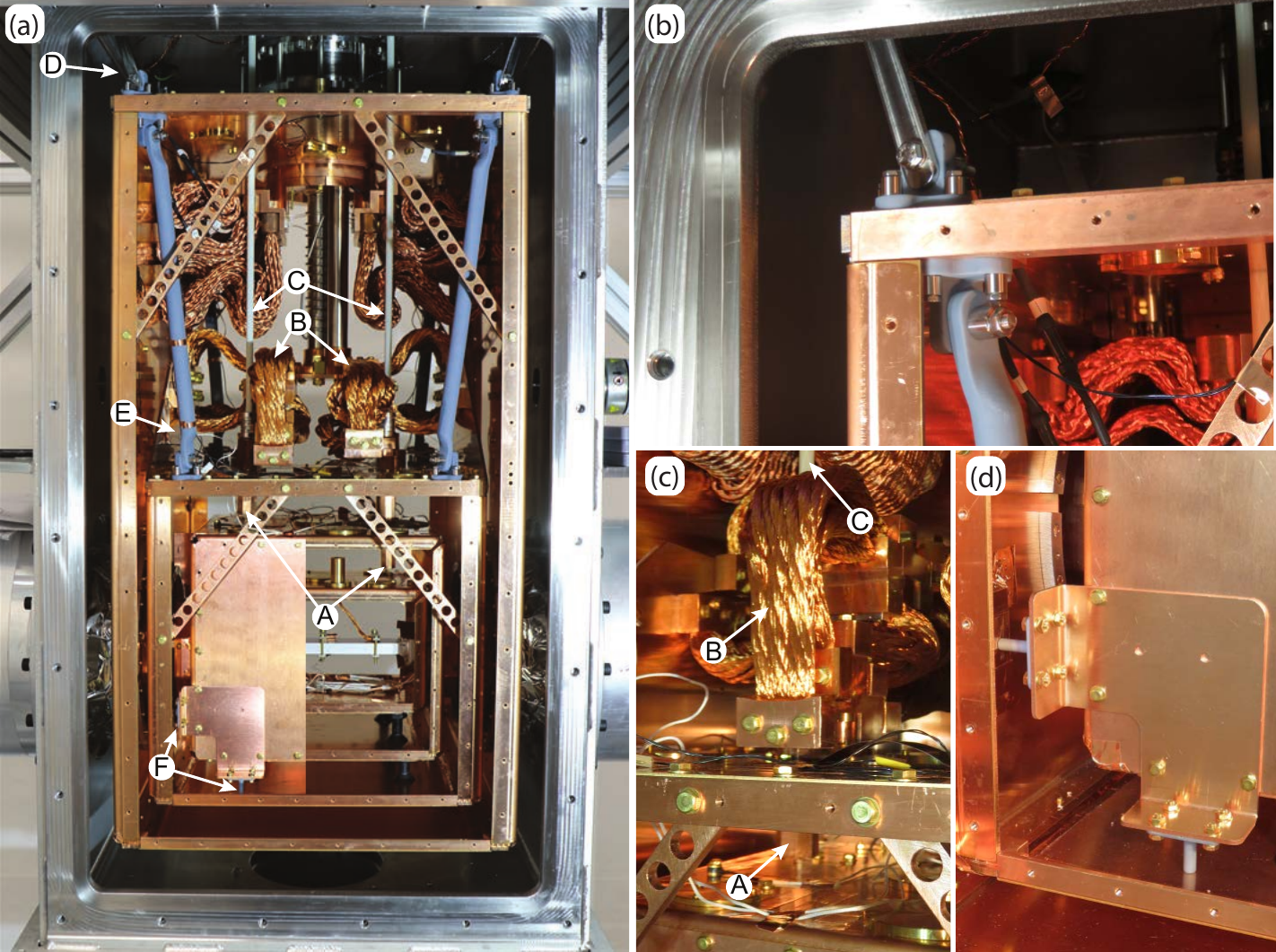}

		\caption{
			(a) Front view of the cooling unit installed at the center of the $30\textrm{\,m}$ link, used to actively cool the 50K and 4K temperature stages. The flanges of the vacuum chamber and radiation shields on the front and back side are removed. Only on the still stage we show the radiation shield on the left half of the picture as a photo composition. The labelled arrows are discussed in the text. (b) Photograph of the support rods for the 50K and 4K stages. (c) Photograph of the still heat switch in the open state (the copper rod is disconnected from the still stage). (d) Photograph of the support posts and spacers of the still stage.
		}
		\label{fig:castor}
	\end{center}
\end{figure*}

The vacuum chamber and the radiation shields, shown in Fig.~\ref{fig:castor} (a), are rectangular, while dilution refrigerators traditionally have cylindrical vacuum chambers and shields. The rectangular shape facilitates access from the front and back side to components mounted at different temperature stages without the need to remove the adapter modules, as is the case for the nodes.

We use copper braids to establish thermal contact between the heat exchange plates at the cold head of the pulse tube and the radiation shields to compensate the thermal contraction of the adjacent elements during the cooldown and decouple the link system from of the vibrations stemming from the pulse tube. 
For optimal heat transport, the cross\nobreakdash-section of the radiation shields along which heat is conducted from the adapter modules to the cold head (including the radiation shield and braids in the cooling unit) is kept constant and similar to the cross\nobreakdash-section of the radiation shields in the link.
We further minimize the number of contact interfaces between the pulse tube and the flanges for the adapter modules and use OFE copper along the main thermal path, resulting in a lower thermal resistance compared to the nodes [see Table~\ref{tab:all_parameters} in Appendix~\ref{thermal_parameters}]. 

When initially cooling down the system starting from room temperature, the pulse tube actively cools the system at the 50K and 4K stages through the braids connecting the stages. To also cool the still and base temperature stages (during the cooldown until condensation of the $^3\textrm{He}/^4\textrm{He}$ mixture), we connect the 4K to the lower temperature stages using mechanical heat switches. 
For our cooling unit, we designed and realized mechanical heat switches, see Fig.~\ref{fig:castor} (a,c), which connect one end of a vertical copper rod to the still or base stage~(A), while the other end is thermalized at the 4K stage through braids~(B). The contact force is generated by an air pressure cylinder mounted outside of the vacuum can and guided to the copper rod through a thermally insulating fiber glass tube~(C). To release the thermal contact of the lower temperature stages to the 4K stage of the pulse tube, for enabling cool down to base temperature, the copper rod is pulled up.
Compared to exchange\nobreakdash-gas\nobreakdash-based heat switches~\cite{Pobell2006}, often used in commercial dilution refrigerators, the mechanical solution has the advantage of not suffering from a residual heat flow between the stages during operation of the cryogenic system at base temperature. However, mechanical heat switches are known to be less reliable than exchange\nobreakdash-gas\nobreakdash-based ones and are therefore rarely used in commercial applications. While a cooldown of the 30-meter-long cryogenic link takes 6.5 days with the mechanical heat switches activated, it increases by a factor of 2 when not using them because the still and base temperature radiation shields along the link are only cooled from the nodes.

The 50K shield is mechanically supported from above by stainless steel rods~(D,b). This standard and reliable approach is suited for the 50K stage as it is the heaviest and in addition is subject to unpredictable lateral forces from the thermal contraction of link elements during the cooldown. 
The 4K stage is mounted to the 50K plate of the cooling unit using 0.3\nobreakdash-m\nobreakdash-long Bluestone rods~(E,b), which are flexible enough to tolerate the typically occurring lateral forces. We use washers made of Invar to  compensate the differences in thermal contraction of Bluestone and stainless steel screws, see Fig.~\ref{fig:castor} (b).

At the still and base stages the space constraints do not allow us to use long and flexible enough Bluestone rods. We mechanically support the still and base stages from below with four Bluestone posts, each mounted at a corner on the bottom of the radiation shields~(F,d), allowing the shields to slide along the link axis for a limited distance of ($ 5\textrm{\,mm}$). The ability for the shields to slide laterally compensates forces caused by the thermal contraction of the still and base\nobreakdash-temperature radiation shields. To avoid accidental thermal shorts, the lateral motion of the still and base stage is limited by the Bluestone spacers mounted horizontally (d).

\section{PRODUCTION AND ASSEMBLY TECHNIQUES}
\label{production_assembly}

In this appendix we discuss three technical aspects of the manufacturing and assembly process of the cryogenic link. The first two, rigid and reliable screw connections under consideration of thermal contraction and checking for radiation leaks are more general while copper braid manufacturing is rather specific for the cryogenic link.

\subsubsection{Achieving rigid screw connections}
Most components in a cryogenic setup are mounted using screws. To ensure good thermalisation of components and rigidity of mechanical connections, we need to choose a proper torque of the screws for their installation. In our cryogenic system, we also have to take into account the different thermal contractions of the screws and the materials they hold together. Otherwise, the screws may break or connections can loosen~\cite{Thomala2007}. Table~\ref{tab:Torque} (2nd column) lists the torques we use for a selection of brass screw sizes connecting copper parts. The values correspond to recommendations for screws with a property class of 4.6~\footnote{Bossard: \href{https://www.bossard.com/global-en/assembly-technology-expert/technical-information-and-tools/technical-resources/preload-and-tightening-torques/approximate-values-for-metric-coarse-threads-vdi-2230/}{Approximate values for metric coarse threads VDI 2230)} (visited on 02.11.2023)}.

It is particularly challenging to connect different segments of the Aluminium  waveguide with each other using brass screws. Since Aluminium with a contraction of $4.15 \textrm{\,mm/m}$~\cite{nist_Al} from room temperature to $<50\textrm{\,K}$ has larger thermal contraction than brass with $3.75 \textrm{\,mm/m}$~\cite{Pobell2006}, the resulting contact pressure is reduced when the waveguide is cooled down. For these connections we have applied the torque for fastening the brass screws displayed in Table~\ref{tab:Torque} (last column), corresponding to the property class of $5.6$. The applied torque results in a preload force at room temperature close to the yield strength of brass. This force is reduced by one quarter at cryogenic temperatures due to thermal contraction of the two materials~\cite{Thomala2007}. 

Even larger mismatches of the thermal contraction coefficients between two materials can be mitigated by adding washers with a low thermal contraction coefficient, e.g. tungsten ($0.75 \textrm{\,mm/m}$) or Invar ($0.5 \textrm{\,mm/m}$)~\cite{Pobell2006}. In the additional cooling unit [see Fig.~\ref{fig:castor}(b) in Appendix~\ref{intermediate_cooler}] we mount the Bluestone rods ($6.5 \textrm{\,mm/m}$)~\cite{Fessia2013} using stainless steel screws ($3 \textrm{\,mm/m}$)~\cite{Pobell2006} and Invar washers.

\begin{table}[t]
\centering
{\footnotesize{}}
    \begin{tabular}{|l | c | c |} 
        \hline
        Thread & \multicolumn{2}{c|}{Torque in Nm for} \\
        size   &\ \ \ \   Cu\ \ \ \  & Al \\
        \hline \hline
        M2.5 & 0.25 & 0.3 \\ 
        \hline
         M3 & 0.40 & 0.5 \\ 
        \hline
         M4 & 0.85 & 1.1 \\ 
        \hline
         M5 & 1.65 & 2.1 \\ 
        \hline
         M6 & 2.8 & 3.6 \\ 
        \hline
        
    \end{tabular}
{\footnotesize{}\caption{\label{tab:Torque}  Our recommended torque to tighten metric brass screws for fastening copper (Cu) or aluminium (Al) parts.}
}
\end{table}

\subsubsection{A method for checking radiation tightness}
Proper shielding of the thermal radiation on every temperature stage is essential to not exceed the cooling power of the colder stages (see main text, Sec.~\ref{sec:design_optimization}). Even with a carefully developed design of the radiation shields, there will always be narrow slits or uncovered holes where radiation can propagate through and warm up the next colder temperature stage. These leaks can be seen as a black\nobreakdash-body surface, increasing the effective emissivity significantly, although their area is small. During the assembly we therefore covered these open holes with copper or aluminium tape to prevent radiation leaking through the stage. An efficient way to find leaks is to place a bright LED light on the inside of the temperature shield, darken the room, and observe where the light can be seen from the outside of the shield. As all the components and shields usually have a high reflectivity in visible wavelengths, the light from the LED will be reflected multiple times inside the shield and even reaches indirectly illuminated areas.

\subsubsection{Manufacturing of copper braids}

A braid module contains several braids at each temperature stage. The total cross\nobreakdash-section of these braids is equal to the cross\nobreakdash-section of the radiation shield of a link module. Each copper braid is an assembly consisting of 2 parts: the braided copper wires, and a terminal (end\nobreakdash-fitting) on each end of the wires. We create a high thermal conductivity connection between the braided wires and the terminal using compression welding. The terminals are then connected to the radiation shields using brass screws.

For braids below the 50K stage we measured the conductivity of the braided copper wires before manufacturing the braids with a similar method as described in Appendix~\ref{RRR_copper}, to select copper with a thermal conductivity $\rho_{\textrm{CU}}\gtrsim 1200 \textrm{\,W/(Km)}$ at $4.2\textrm{\,K}$. During the manufacturing process, thermal conductivity of a copper braid can be lowered due to repeated bending of the braided copper wires, and the plastic deformation while compression welding the terminals. In a final production step, we anneal the braids in a vacuum oven~\cite{Pobell2006}, aiming to restore or increase the thermal conductivity of the braids.

\section{THERMAL MODEL}
\label{thermal_model}

We developed a simplified thermal model in parallel to the development of the cryogenic link and prototype setups in an iterative approach, with the aim to both optimize the design of the cryogenic link, and to improve the understanding of the heat transfer processes in the system. In the following, we consider in some detail the steady\nobreakdash-state heat transfer across different stages of our system, and discuss the two dominant heat\nobreakdash-transfer mechanisms acting on the individual stages.

We label the base, still, 4K, 50K and room temperature stages (vacuum can) with $n = 1 ... 5$, and consider a general cryogenic link with length $l$. On each stage $n$, we assume that the heat load originates exclusively from the enclosing higher temperature stage $n+1$, [see Fig.~\ref{fig:thermal_model}(a)]. The temperature of the outermost shield $n=5$ (vacuum can), is considered to be constant at room temperature.

We model each temperature stage as a one\nobreakdash-dimensional object subjected to heat loads that vary only along the link axis. Because the radiative heat load is cylindrically symmetric, and the thermal conduction of the support posts is several orders of magnitude smaller than that of copper, the heat introduced by the posts rapidly redistributes around the radiation shield. We therefore neglect any circumferential temperature variation of the shield induced by the posts.

\subsubsection{Heat load}
We first consider thermal radiation and local heat load induced by the conduction through the posts, and present the expressions used to evaluate those quantities at every position along the link.

The radiated power of a body per unit area can be described with the Stefan–Boltzmann law as $\varphi_{\textrm{rad}}= \sigma \varepsilon T^4$ with the Stefan–Boltzmann constant $\sigma$ and the temperature of the body $T$. The emissivity $\varepsilon$ of the surface of the body describes the efficiency in emitting thermal radiation, being 1 for an ideal black body and $ \varepsilon \in \left ( 0,1 \right )$ for real materials.

We approximate the radiative heat load per unit area on the octagonal shield from stage $n+1$ to stage $n$ by considering two concentric cylindrical grey bodies with emissivity $ \varepsilon_n,\varepsilon_{n+1}\ll1 $ by

\begin{equation}
    \varphi_{\textrm{rad},n}\simeq \sigma\lambda T_{n+1}^4,
    \label{eq:radiation}
\end{equation}
with the adapted attenuation factor $\lambda \simeq (\frac{1}{\epsilon_n}+\frac{F}{\epsilon_{n+1}})^{-1}$~\cite{Parma2015,Lienhard2020}. This includes the view factor $F=C_n/C_{n+1}$ with the circumference $C_n$ of the octagonal radiation shield at stage $n$, which we approximate with the average of the circumcircle and incircle [see Table~\ref{tab:all_parameters} in Appendix~\ref{thermal_parameters} for specific values]. $F$ describes the proportion of the total radiation emitted from the outer surface $n+1$ that is absorbed by the enclosed shield $n$. We also neglect the radiation from stage $n$ to $n+1$ since $T_n^4 \ll T_{n+1}^4$.

The second major contribution to the heat load is the conductive heat flow through the support posts. Other elements connecting different temperature stages, such as electronic wiring, only appear in the nodes and are designed to minimize thermal conduction and are therefore not considered in our thermal model.

Heat transferred inside an object through thermal conductance can be described with the local heat flux density $\vec{q}$, which is governed by Fourier's law~\cite{Lienhard2020}. In its differential form,

\begin{equation}
    \vec{q} = -\rho(T) \nabla T, 
    \label{eq:fouriers_law}
\end{equation}
where $\rho(T)$ is the temperature-dependent thermal conductivity of the material of the object, and $\nabla T$ is the temperature gradient within the object.

Each post in the link makes physical contact with two adjacent temperature stages, inducing heat flow from the hotter to the colder end of the post. We model the heat transferred through a post as a one\nobreakdash-dimensional process, since the cross\nobreakdash-section of a post is approximately constant, reducing Eq.~\eqref{eq:fouriers_law} to $q_x=-\rho dT/dx$. Next, we put $dx$ to the left-hand side and multiply by the average cross-sectional area $A_{\textrm{post},n}$. Integrating both sides results in the heat load per post on stage $n$ as

\begin{align}
P_{\textrm{post},n} &= \frac{A_{\textrm{post},n}}{l_{\textrm{post},n}}  \int_{T_{n}}^{T_{n+1}}{\rho_{\textrm{post}}(T)} dT \label{eq:conduction0} \\
       &\simeq \frac{A_{\textrm{post},n}}{l_{\textrm{post},n}}  \int_0^{T_{n+1}}{\rho_{\textrm{post}}(T)} dT,\label{eq:conduction}
\end{align}

$l_{\textrm{post},n}$ denotes the length of the post. We approximate the lower temperature as zero in Eq.~\eqref{eq:conduction}, which is valid for common materials with thermal conductivity monotonically decreasing towards low temperatures and under $T_n \ll T_{n+1}$ (see
Appendix~\ref{thermal_properties_bluestone} and Ref~\cite{Pobell2006}). Here, we neglect the radiation heat load on the post itself, assuming it to be small. The estimated values for the thermal model are summarized in Table~\ref{tab:all_parameters} in Appendix~\ref{thermal_parameters}. 

For a given stage $n$, the heat loads in Eqs.~\eqref{eq:radiation}~and~\eqref{eq:conduction} are  determined by the geometrical design, material properties and the temperature profile $T_{n+1}(x)$ of stage $n+1$.
The heat exchange between the temperature stage $n+1$ and $n$ dominates over the exchange between $n$ and $n-1$, because the heat load in Eqs.~\eqref{eq:radiation}~and~\eqref{eq:conduction} scales at least linearly in $T$, and the temperature stages in a dilution refrigerator follow $T_{n-1} \ll T_n \ll T_{n+1}$. This allows approximating the heat load at every position $x$ of stage $n$ in a cascaded manner without considering the colder stage $n-1$.

\subsubsection{Heat flow}
Now we discuss how the heat flow along the link towards a cooling unit is related to the material and geometry of the radiation shields.

Assuming the cooling units at each end of the system to provide equal cooling power makes the model symmetric around the middle of the link, allowing for a simplified study of only one half of the system. All the heat load of this half must therefore flow towards the corresponding cooling node as shown in Fig.~\ref{fig:thermal_model}(c). In combination, the heat load along the radiation shield at stage $n$ and at position $x<l/2$ for steady state condition can be described as a sum of continuous\nobreakdash-in\nobreakdash-space radiative and discrete\nobreakdash-in\nobreakdash-space conductive components:

\begin{equation}
    Q_n(x)  :=\int_{x}^{l/2}\varphi_{\textrm{rad},n}(x')C_n dx'+ \sum_{x_i \geq  x} P_{\textrm{post},n,x_i}.
    \label{eq:heat_current}
\end{equation}
The first term describes the radiative heat load $\varphi_{\textrm{rad},n}$ integrated from the position of interest $x$ to $l/2$ times the circumference $C_n$ of the shield, and the second term describes the conductive heat load from support posts. We denote the position of the posts by $x_i$, and only consider the posts located between $x$ and $l/2$.

The cross-section of radiation shields and braid modules is designed to be constant along the link for each temperature stage, and the geometry is cylindrically symmetric. Therefore, we again use the one\nobreakdash-dimensional Fourier law. At a position $x$ on stage $n$ we relate the temperature gradient $dT_n(x)/dx$ along the link axis to the heat load of Eq.~\eqref{eq:heat_current} as follows:

\begin{equation}
    A_n \rho_n(T_n(x))\frac{dT_n(x)}{dx}=Q_n(x),
    \label{eq:temp_profile}
\end{equation}
with the temperature\nobreakdash-dependent thermal conductivity of the radiation shield $\rho_n(T)$, its cross\nobreakdash-sectional area $A_n$, the radiation shield temperature $T_{n}(x)$ at the position of interest. The parameters $A_n$ and $\rho_n$ are listed in Table~\ref{tab:all_parameters} in Appendix~\ref{thermal_parameters} .

In addition, we model the thermal interface between two bulk materials such as between a braid module and a radiation shield using a temperature\nobreakdash-dependent contact resistance $R_{\textrm{c}}(T)$. Correspondingly, a heat flow crossing such an interface leads to a finite temperature jump~\cite{Salerno1997} 
\begin{equation}
    \Delta T =R_{\textrm{contact}}(T) Q.
    \label{eq:contact_res}
\end{equation}
The contact resistance for braid modules is discussed in more details in Appendix~\ref{thermal_properties_link}.

\subsubsection{Temperature profile}

\begin{figure}[t]
	\begin{center}
		\includegraphics[width=1.0\columnwidth]{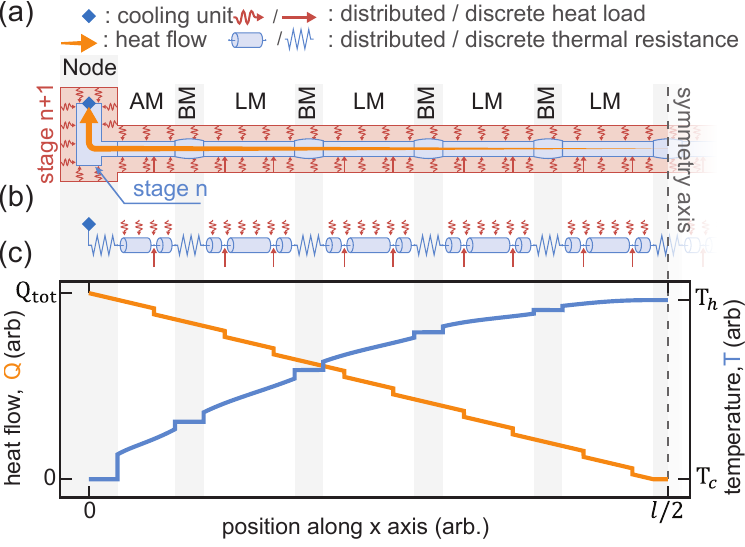}
		\caption{
			(a) Schematic representation of heat transfer from stage $n + 1$ to the next colder stage $n$. The abbreviations AM, BM and LM stand for adapter module, braid module and link module. The increasing width of the orange arrow depicts the increasing heat flow along the link. (b) Equivalent thermal circuit model of (a). (c) Representative curves of the heat flow (orange) and temperature (blue) profiles along the link. The total induced heat flow $Q_{tot}$ is the sum of all heat loads on the half section of stage n, $T_c$ is the temperature at the cooling unit and $T_h$ is the maximum temperature.
		}
		\label{fig:thermal_model}
	\end{center}
\end{figure}

The simplified model outlined above allows us to numerically calculate the temperature profile of every stage along the link. We begin with the vacuum can, which is in good approximation at room temperature along its full length. We calculate the temperature profile $T_n(x)$ of any lower temperature stage $n$ using the following steps.
\begin{enumerate}

    \item \textbf{Calculate $ \bm{Q_n(x)}$:} By taking into account the dimensions and material properties of the radiation shields and support posts [see Table~\ref{tab:all_parameters} in Appendix~\ref{thermal_parameters}], we calculate the heat flow rate incident on stage $n$ using Eq.~\eqref{eq:heat_current}, given a known temperature profile $T_{n+1}$ at stage $n+1$. In this approach, we model the link modules as a set of connected elements (as shown in Fig.~\ref{fig:thermal_model}(b)). The shield and connection modules are considered as distributed thermal resistances with discrete heat load stemming from the posts and distributed heat load due to the incident thermal radiation. The nodes and braid modules are considered as thermal resistances causing a finite temperature jump at their interfaces. By integrating Eq.~\eqref{eq:heat_current}, we calculate the heat flow along the link $Q_n(x)$. 
    
    \item \textbf{Determine the node temperature $ \bm{T_n(0)}$:} At the cooling unit ($x=0$), the total heat flow equates to the temperature\nobreakdash-dependent cooling power of the unit given by the relation $T_n(0) = P_n [Q_n(0)]$, with $P_n$ denoting the cooling power curve of the node at stage n [see Table~\ref{tab:all_parameters} in Appendix~\ref{thermal_parameters}].
    
    \item \textbf{Calculate $ \bm{T_n(x)}$:} Let ${x_i}$ be the ordered finite set of $M$ locations with discrete thermal resistance modules: $0 \leq x_1 \leq ... < x_M \leq l/2$. For a given thermal contact at location $x_i$, we define $T_n (x_{i-})$ and $T_n (x_{i+})$ as the temperatures just before and after the thermal contact. We then iteratively calculate the temperature profiles. If $x_1>0$, we integrate Eq.~\eqref{eq:temp_profile} from $0$ to $x_{1-}$ to get $T_n(x)$ on this section of the link and $T_n(x_{1-})$. Then for $i = 1 \textrm{ to } M$: 
    \begin{enumerate}
        \item we solve Eq.~\eqref{eq:contact_res} to find $T(x_{i+})$ using $T(x_{i+}) - T(x_{i-}) = R_{\textrm{module}}[ {T(x_{i+}) + T(x_{i-}) }/2 ] Q_n(x_i)$.
        \item  and integrate Eq.~\eqref{eq:temp_profile} from $x_{i+} \textrm{ to } x_{(i+1)-}$.
    \end{enumerate}

\end{enumerate}

These steps are repeated in a cascaded manner until we find the profile of the lowest temperature stage.

An example profile of $Q_n$ as a function of distance $x$ from the node is shown in Fig.~\ref{fig:thermal_model}(c). Since all heat loads from the posts and radiation are positive, $Q_n(x)$ increases continuously from $x=l/2$ towards the node at $x = 0$ with steps at the positions where the support posts are installed on the radiation shields. 
The temperature gradient $  dT_n/dx$ is approximately proportional to the heat flow. It reaches the maximum at the node. In the middle of the link $l/2$ the heat flow and therefore the temperature gradient vanishes. Discrete temperature jumps occur for modules modelled as discrete thermal resistance.

\begin{figure*}[ht]
	\begin{center}
		\includegraphics[width=2.0\columnwidth]{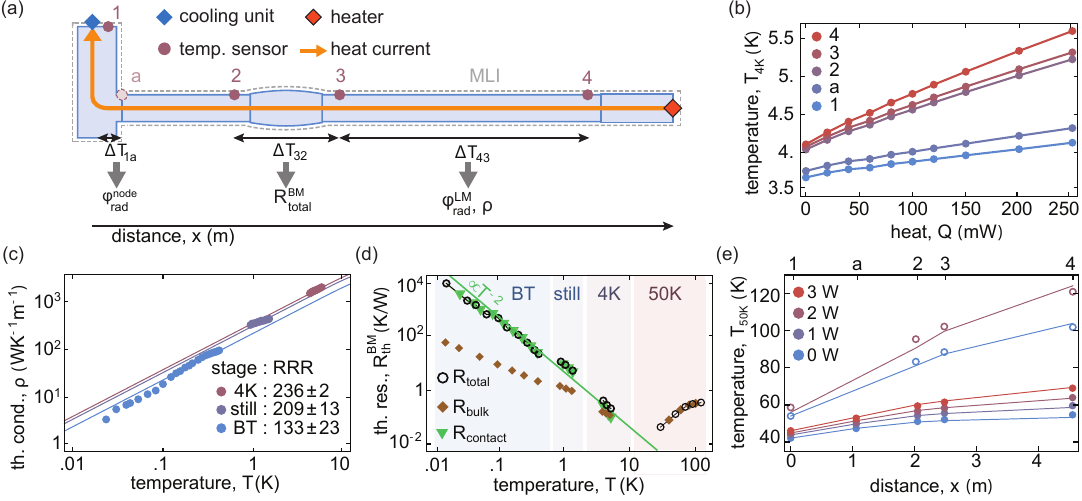}
		\caption{
		(a) Scheme of the experimental setup used to characterize heat transfer in the link and braid modules. (b) 4K stage temperature distribution {\it vs} heat applied at the end of the link. (c) Temperature dependence of the thermal conductivity at the base temperature, still and 4K shields of the link module. The solid lines are fits of the data to the RRR model (see Appendix~\ref{RRR_copper}). (d) Temperature dependence of the thermal resistances of the braid module (BM) and its contributions from finite bulk conductivity and contact resistances. The green line highlights the approximate $ T^{-2} $ dependence of interface resistances. (e) 50K steady-state temperature distribution in the presence (filled circles) or absence (empty circles) of multi-layer insulation, for the indicated applied heat power. The solid lines are results of finite elements simulation, with the heat flux as a free parameter. The distance $x$ also takes the vertical path of the heat flow to the cooling unit into account.
		}
		\label{fig:thermal_properties_modules}
	\end{center}
\end{figure*}

\section{THERMAL PROPERTIES OF LINK AND BRAID MODULES}
\label{thermal_properties_link}

In this section, we discuss the characterization of the thermal conductivity of the radiation shields, the thermal resistance of braid modules and the effect of multi\nobreakdash-layer insulation on the 50K stage. The results of this analysis serve as parameters for the thermal model (see Table~\ref{tab:all_parameters} in Appendix~\ref{thermal_parameters}), which is used to simulate the temperature profile of the cryogenic link cooldowns and to predict the achievable lengths of the cryogenic link discussed in the main text (Sec.~\ref{sec:cooldown_results}). For the characterization of these properties we assembled a cryogenic prototype system consisting of a single node connected to an adapter module and a link module, as depicted in Fig.~\ref{fig:thermal_properties_modules}(a). At the end of the only link module of the system, we closed the radiation shields with radiation\nobreakdash-tight caps. At each temperature stage we mounted temperature sensors at the positions indicated by the purple dots in Fig.~\ref{fig:thermal_properties_modules}(a) and a heater at the end of the link on the cap (red diamond). 

\subsubsection{Thermal conductivity of the radiation shields}

First, we experimentally evaluate the thermal conductivity of the radiation shields in a link module. For this purpose we measure the temperature profile of the cold prototype system at steady state when applying different powers $Q$ with the heater at the end of the link module for each temperature stage separately [Fig.~\ref{fig:thermal_properties_modules}(a)]. 
We measure the temperatures at four different location, across all temperature stages: Sensor 1 is placed on the temperature plates of the dilution refrigerator, sensor 2 at the end of the adapter module, and sensors 3 and 4 at each end of the radiation shield of the link module.

Figure~\ref{fig:thermal_properties_modules}(b) shows the temperatures measured at the 4K stage versus the applied heating power. We observe that in the absence of additional heating, the temperatures increase along the link due to the heat load of the support posts and thermal radiation in the system (see Appendix~\ref{thermal_model}). When increasing the applied heat $Q$, the temperature difference between neighbouring sensors increases due to the finite thermal conductivity of the radiation shields and braids.

We use these measurement results to calculate the thermal conductivity of the radiation shields from the extracted temperatures of sensors 3 and 4 {\it vs} applied power.
For this purpose, as discussed in the thermal model (Appendix~\ref{thermal_model}), we approximate the geometry of a radiation shield for the heat transfer between sensors 3 and 4 as a one\nobreakdash-dimensional conductor. The differential form of the Fourier's law then reads 

\begin{equation}
    Q_n+Q_{n,0}(x) =   A_n \rho^{\textrm{LM}}_n(T_n(x))\frac{\partial T_n}{\partial x},
    \label{eq:temp_profile_shield}
\end{equation}
with $n$ denoting the temperature stage, $A_n$ the cross section area of the radiation shield, $\rho^{\textrm{LM}}_n$ its thermal conductivity, $T_n(x)$ the temperature profile along the shield and $Q_{n,0}(x)$ the sum of all heat loads at temperature stage $n$ between position $x$ and the end of the prototype link. The heat load from the heater is represented by $Q_n$. Next, we integrate Eq.~\eqref{eq:temp_profile_shield} between sensor 3 and 4 using a first order approximation of $\rho^{\textrm{LM}}_n$ around the mean temperature $\Bar{T}_{n,34}:=(T_{n,3}+T_{n,4})/2$. We find the following expression for the thermal conductivity of the shield

\begin{equation}
    \rho^{\textrm{LM}}_n(\Bar{T}_{n,34}) \simeq \frac{\Delta x_{n,43} (Q_n+\Bar{Q}_{n,34})}{A_n \Delta T_{n,43}}, 
    \label{eq:temp_profile_shield_approx}
\end{equation}
where $\Bar{Q}_{n,34}$ is the mean value of the heat load $Q_{n,0}(x)$, $\Delta x_{n,43}$ the distance and $\Delta T_{n,43}:=T_{n,4}-T_{n,3}$ the temperature difference between sensor 3 and 4.

With Eq.~\eqref{eq:temp_profile_shield_approx} and the temperatures measured for different heating powers, we calculate the thermal conductivity $\rho^{\textrm{LM}}_n$ of the radiation shield at stage $n$. The results for the base, still and 4K radiation shields are displayed in Fig.~\ref{fig:thermal_properties_modules}(c). The points represent measurement data and the colored lines are fits of the residual resistivity ratio (RRR) model, by minimizing the quadratic error to the measured thermal conductivity. This model uses a single parameter to describe the thermal conductivity of copper (see Appendix~\ref{RRR_copper}). The analysis for temperatures above 50~K is particularly challenging, as the thermal conductivity $\rho(T)$ of the different RRR values in the model only differs by a few percent [see Fig.~\ref{fig:RRR_measurement}(d)]. For this reason we did not measure it.

The limited temperature range for the heated radiation shields results in an increased RRR variability with a standard deviation of up to 20\% of the mean, see legend in Fig.~\ref{fig:thermal_properties_modules} (c). However, the RRR values are similar to the RRR values used for the thermal model (see Sec.~\ref{thermal_parameters}), which are based on measurements of test strips cut from the copper sheets prior to their use in manufacturing the radiation shields. The measurement method itself is discussed in Appendix~\ref{RRR_copper}. This indicates that the mechanical processing, such as laser cutting and bending used in the manufacturing process of the radiation shields, does not significantly affect the observed thermal conductivity of the copper~\cite{Pobell2006}.

\subsubsection{Thermal resistance of the braid modules}

Next, we aim to understand the fraction of the thermal resistance along the braid module stemming from the limited thermal conductivity of the braids themselves and from the contact resistance between the braids and radiation shields.

For this purpose, we define the total thermal resistance $R_{\textrm{total}}^{\textrm{BM}}$ of the braid module as the inverse of the thermal conductance. In combination with its cross\nobreakdash-section area $A_n$ and length $\Delta x_{n,32}$ accounting for the geometry of the braid module we can approximate the thermal resistance using similar considerations as in Eq.~\eqref{eq:temp_profile_shield_approx}:

\begin{equation}
    R_{\textrm{total},n}^{\textrm{BM}}(\Bar{T})  \simeq \frac{\Delta T_{n,32}}{Q_n+ \Bar{Q}_{n,23}},
    \label{eq:temp_R_th_shield}
\end{equation}

with measured temperature difference $\Delta T_{n,32}$, the applied heat $Q_n$ and the mean heat current of the braid module $\Bar{Q}_{n,23}$. We assume that the latter is similar to the mean heat current of the link module ($ \Bar{Q}_{n,23} \approx \Bar{Q}_{n,34}$) as there is no support post inside the braid module that would introduce extra conductive heat load and as the radiative heat load inside a braid module can be neglected because its surface area is less than $1/10$ of the surface area of the radiation shield.

The measured total thermal resistance is represented in Fig.~\ref{fig:thermal_properties_modules}(d) with black rings. Due to the thermal conductivity properties of copper and $R_{\textrm{total}}^{\textrm{BM}}(T) \propto \rho(T)^{-1}$ the resistivity reaches its lowest point around $20 \textrm{\,K}$.

To differentiate between thermal resistance stemming from finite thermal conductivity of the braids and from contact resistance between the braids and the radiation shields, we write the total thermal resistance as a combination of a bulk part and a contact resistance, $ R_{\textrm{total}}^{\textrm{BM}}(T) = R_{\textrm{contact}}^{\textrm{BM}}(T) + R_{\textrm{bulk}}^{\textrm{BM}}(T)$.

We determine the bulk thermal resistance $ R_{\textrm{bulk}}^{\textrm{BM}} $ of the braids, which are used in the braid modules, in a dedicated measurement using a similar method as discussed in Appendix~\ref{RRR_copper} for the copper test strips. The results are shown in Fig.~\ref{fig:thermal_properties_modules}(d) with brown diamonds. This allows to extract the contact resistance by combining both measurements,
$R_{\textrm{contact}}^{\textrm{BM}} =R_{\textrm{total}}^{\textrm{BM}} - R_{\textrm{bulk}}^{\textrm{BM}}$, represented with green triangles in the figure. The measured resistances are proportional to $T^{-2}$ below 10~K (green line in Fig.~\ref{fig:thermal_properties_modules}(d)). This indicates that at low temperatures, such as on the base and still stage, contact resistance dominates. On the 4K stage the contribution of contact and bulk resistance are approximately equal, and at the 50K stage we observe that the contact resistance is negligibly low.

It is generally difficult to make theoretical predictions about the contact resistance between bulk materials, and the problem is not well-understood~\cite{Salerno1997,Pobell2006}. However, at low temperatures the contact resistance typically follows the relation $R_{\textrm{contact}} \propto T^{-b},\, b \in [0.75,2.5]$~\cite{Salerno1997}. Oxide layers on the surfaces, which act as an additional boundary resistance, introduce an acoustic mismatch between the layers (Kapitza resistance)~\cite{Salerno1997} and affect the observed temperature dependence $b$. General advices for lowering contact resistances at low temperatures $\leq 4\textrm{\,K}$ are to use smooth electro\nobreakdash-polished copper surfaces free of oxides and pressing them together with high force. A thin film of thermal conducting vacuum grease like Apiezon N applied to the interface can reduce the contact resistance further~\cite{apiezon}, because the grease fills the tiny crevices on the surface and increases the effective contact area to transfer heat.

\subsubsection{Multi-layer insulation}

In the last step we study the impact of multi\nobreakdash-layer insulation installed on the 50K stage of the link. For this purpose, we cooled down the prototype system (see Fig.~\ref{fig:thermal_properties_modules}(a)) twice: once with the 50K stage directly exposed to the room-temperature radiation from the vacuum can, and the second time with multi-layer insulation (MLI) covering the 50K radiation shields (indicated by a dashed line in Fig.~\ref{fig:thermal_properties_modules}(a)).

The measured temperature profiles for the two consecutive cool downs are shown in  Fig.~\ref{fig:thermal_properties_modules}(e). Without the MLI (empty circles), the temperature gradient along the prototype link is larger than with MLI (filled circles) and reaches $100\textrm{\,K}$/$120\textrm{\,K}$ at the end of the shield module without/with heater switched on. With the MLI installed, the temperature is lower than without, everywhere and for all applied heating powers.

For the analysis of the 50K stage we use a 3D model of the prototype system and a finite element simulation software. We assume that the mean heat flux $\varphi_{\textrm{eff}}$ on the 50K shield scales as $\varphi_{\textrm{eff}}=\lambda\sigma(T^4_{\textrm{RT}}-T(x)^4)$, where $\sigma$ is the Stefan\nobreakdash-Boltzmann constant, $\lambda \in [0,1]$ the attenuation coefficient composed of the surface emissivities and a geometric factor (see Eq.~\eqref{eq:radiation}).

First, we analyze the system without multi-layer insulation installed. We simulate the steady\nobreakdash-state temperature profile by sweeping $\lambda$, and find the minimum quadratic error between the measured and simulated temperatures for $\lambda=2 \%$, see lines in Fig.~\ref{fig:thermal_properties_modules}(e). This value corresponds to a  mean heat flux of $\varphi_{ \textrm{eff}}=8.5 \textrm{\,W/m}^2$. The mean heat flux in the simulation is a combination of the thermal irradiation $\varphi_{\textrm{rad}}$ from the vacuum can and the heat load of a single post $P_{\textrm{post}}$ times the mean number of posts per unit area $n_p$ separating the 50K shield from the vacuum can: $ \varphi_{\textrm{eff}} = \varphi_{\textrm{rad}}+n_p\, P_{\textrm{post}} $. From the characterization and simulation of the support posts we expect that each post on the 50K stage adds a heat load of $P_{\textrm{post}} \approx 0.8\textrm{\,W}$ (see Appendix~\ref{thermal_properties_bluestone}). In the prototype system the post density is $n_p\approx 2.6\textrm{\,m}^{-2}$. Subtracting the heat load stemming from the support posts from $\varphi_{\textrm{eff}}$ leads to an approximate radiation heat load of $\varphi_{\textrm{rad}}\approx 6.4\textrm{\,W/m}^2$. 

Second, we analyze the system with multi-layer insulation covering the 50K radiation shield of the link modules. The MLI consists of two custom\nobreakdash-designed sets with 15 layers each. On the node the only one set of MLI is installed due to space constraints between the vacuum can and the 50K shields. To distinguish the performance of the MLI on the link modules from the performance on the node, we additionally install a temperature sensor along the 50K shield of the dilution refrigerator, labelled \textit{a} in Fig.~\ref{fig:thermal_properties_modules}(a). In the simulation we split the attenuation constant $\lambda$ into two parts: $\lambda_{1a}$ represents the section of the radiation shield of the node, and $\lambda'$ the section of the radiation shields along the link.

Again, we simulate the steady\nobreakdash-state temperatures and find a minimum quadratic error between measurement and simulation data for $\lambda' = 0.4 \% $, corresponding to $\varphi_{\textrm{eff}}=1.7 \textrm{\,W/m}^2$. The heat load of a post, which is now partially covered with MLI, is estimated to be $P_{\textrm{post}} \approx 0.26\textrm{\,W}$ (see Appendix~\ref{thermal_properties_bluestone}). For the radiation heat load we find $\varphi_{\textrm{rad}}\approx 1.0\textrm{\,W/m}^2$, which is comparable to other studies using similar MLI configurations~\cite{Shu1986,Fesmire2002,Parma2015}. The MLI significantly reduces the radiation heat load on the 50K stage and is therefore a critical element for the successful operation of the system at length scales of tens of meters.

For the node section ($\lambda_{1a}$) the MLI did not substantially improve the reflectivity. We attribute this to the surface geometry of the radiation shield and tight space constraints between the radiation shield and vacuum can, which makes it difficult to properly attach the MLI and fully cover the whole surface. Open ends of the MLI can also behave as black body absorbers~\cite{Parma2015, marlow_design_2011}. We decided not to cover the nodes with MLI for the cryogenic link.

\section{CRYOGENIC PROPERTIES OF BLUESTONE} 
\label{thermal_properties_bluestone}

\begin{figure}[b]
	\begin{center}
		\includegraphics[width=1.0\columnwidth]{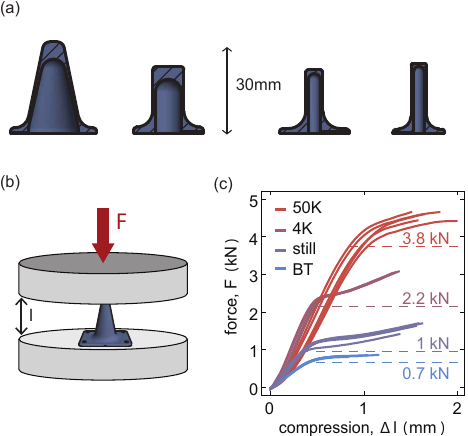}
		\caption{
			(a) Cross-sectional CAD drawings of the 50K (left) to base-temperature (right) Bluestone posts. (b) Schematic representation of the compression test machine used to characterize the compressive yield force of the posts. (c) Controlled force applied on the post {\it vs} measured post compression for several copies of each type of post. The dashed line indicate the mean elastic limit after which permanent deformation of $ 0.2\% $ of the length $l$ occurs.
		}
		\label{fig:bluestone}
	\end{center}
\end{figure}

In this section, we discuss mechanical and thermal properties of different material candidates. We consider Macor, polyether ether ketone (PEEK), glass- and carbon\nobreakdash-fiber\nobreakdash-reinforced polyamid (GF/CF Nylon) and Bluestone~\cite{accura_bluestone} for the mechanical support posts.

\subsubsection{Yield strength}
To analyze the suitability of Bluestone as a material of choice to mechanically support the shields in the link modules, we test the yield strength of our custom thin\nobreakdash-walled posts at room temperature, see Fig.~\ref{fig:bluestone} (a), with a standard compression testing machine~\cite{alma99}. A post is placed between a lower and an upper compression plates, schematically shown in Fig.~\ref{fig:bluestone}(b). The distance $l$ between the plates is constantly measured while the vertical force $F$ on the post is increased. The initial lengths of the posts $l_0$, for the four different temperature stages, when no force is applied lie between $20\textrm{\,mm}$ and $32\textrm{\,mm}$. For low forces, the compression $\Delta l=l_0-l$ is proportional to $F$ due to elastic deformation of the material, see straight segment of the lines in Fig.~\ref{fig:bluestone}(c). Increasing the force further leads to a plastic deformation of the material (flattening of the lines) and later mechanical rupture (end point of the lines).

We characterize the effective compressive strength of the material with the offset yield strength $R_{\textrm{p}0.2}$~\cite{Beer2012}, defined as the amount of stress that leads to a plastic deformation of $0.2\%$. The five tests on the 50K posts yield an $R_{\textrm{p}0.2}$ variation of 5\%. This small variation between measured yield strengths shows that the production of posts based on a 3D printing process is reliable. 
Due to our choice of measuring the five different post geometries, and not a typical geometry for compression testing (e.g. a cylinder or cuboid), we obtain a larger variation $R_{\textrm{p}0.2}$ in a range from $100$ to $130\textrm{\,MPa}$.

We also performed the same tests for posts made from Macor, PEEK and CF/GF Nylon, see Table~\ref{tab:bluestone_summary}. For Stainless steel we reference values from the literature~\cite{Marquardt2002}.

\subsubsection{Thermal conductivity}
\begin{figure}
	\begin{center}
		\includegraphics[width=1.0\columnwidth]{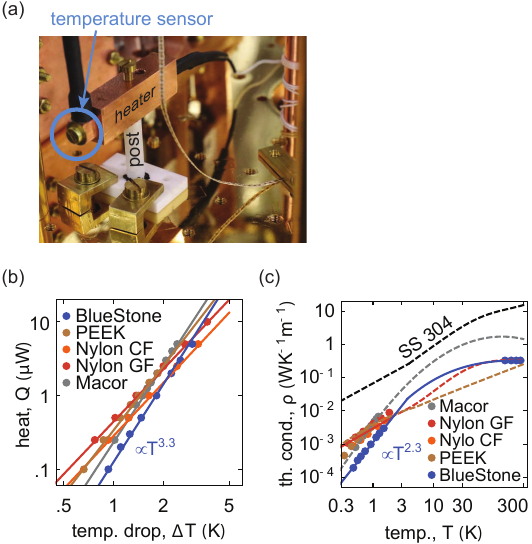}
		\caption{
			(a) Photograph of a typical experimental setup used to characterize the conductive heat properties of base temperature posts made of Bluestone, polyether ether ketone, Nylon and Macor. The post is thermalized to the base temperature plate of a dilution refrigerator. (b) Applied heat {\it vs} measured temperature drop across the posts from (a). The solid lines are power law fits (see main text). (c) Thermal conductivity extracted from (b). The solid line for Bluestone is obtained as a piecewise function which fits the data at different temperature scales. The dashed lines are values taken from literature~\cite{Marquardt2002,Woodcraft2009}. 
		}
		\label{fig:bluestone2}
	\end{center}
\end{figure}

Mechanical support posts contribute to the conductive heat load exerted on the temperature stages (see Appendix~\ref{thermal_model}). We are not aware of data on the thermal conductivity of Bluestone at cryogenic temperatures suitable for us, so we have characterized the Bluestone posts under these conditions. 
For these measurements, we clamp the base of a post to the base temperature plate of a dilution refrigerator. A heater and a resistive ruthenium oxide temperature sensor (RX-102B from LakeShore Cryotronics, Inc.) are attached to a copper block with screws, and this block is glued to the top head of the post as shown in Fig.~\ref{fig:bluestone2}(a) for a Macor post. We use the temperature sensor to monitor the temperature $T_{\textrm{top}}$. The sensor is electrically connected using a superconducting NbTi core enclosed in a low-conductivity CuNi matrix, ensuring that the heat flow through the wire is negligible compared to the heat flow through the post. A second temperature sensor is mounted on the base plate to measure the temperature of the base of the post $T_{\textrm{base}}$  (not visible on the photograph). 

The temperature difference between the two sensors $\Delta T = T_{\mathrm{top}}-T_{\mathrm{base}}$, measured while applying heating powers $Q$ ranging from $0.1 \,\mu\textrm{W}$ to $10 \,\mu\textrm{W}$ for posts with the same geometries but made of different materials, is shown in Fig.~\ref{fig:bluestone2}(b). During measurements, $T_{\textrm{base}}$ varied only from $10\textrm{\,mK}$ to $20\textrm{\,mK}$, while $T_{\textrm{top}}$ peaked at $10\textrm{\,K}$.  Compared to the other tested materials, Bluestone shows the steepest temperature drop ($\propto \Delta T^{3.3}$) below $5\textrm{\,K}$. For $\Delta T < 1.5\textrm{\,K}$ the heat current flowing through the Bluestone post appears to be only half as large as for any other material tested, while above $\Delta T > 4\textrm{\,K}$, Nylon and PEEK are better thermal insulators. We fit the data by minimizing the quadratic error to the measured data using a power law model $Q=a \Delta T^b$. For Bluestone we find a dependence scaling as $\propto \Delta T^{3.3}$ with the temperature difference [see Fig.~\ref{fig:bluestone2}(b)].

To calculate the thermal conductivity for each material, we use the heat load calculation of the posts from the thermal model (see Eq. \eqref{eq:conduction}), and approximate $T_{\mathrm{base}}=0$~K since $T_{\textrm{base}} \ll T_{\textrm{top}} $. We get the relation $\Delta T = T_{\textrm{top}} $ and rewrite Eq.~\eqref{eq:conduction} as

\begin{equation}
    Q\frac{l_{\textrm{post}}}{A_{\textrm{post}}}= \int_{0}^{T_{\textrm{top}}} \rho_{\textrm{post}}(T)dT, 
    \label{eq:conduction_post}
\end{equation}
with the length $l_{\textrm{post}}$ and the average cross\nobreakdash-section $A_{\textrm{post}}$ of the post. As an ansatz we write the heat current as $Q(T) =aT^b$ and differentiate Eq.~\eqref{eq:conduction_post} with respect to T, which gives us the thermal conductivity as $\rho_{\textrm{post}}(T)=abT^{b-1} l_{\textrm{post}}/A_{\textrm{post}}$. Below $4\textrm{\,K}$ we estimate from the experiment that the thermal conductivity of Bluestone scales with $T^{2.3}$ [see Fig.~\ref{fig:bluestone2}(c)].

For the thermal conductivity between $50\textrm{\,K}$ and $300 \textrm{\,K}$, we repeated the same measurements by mounting a 50K post on the 50K stage inside the corresponding radiation shield. In this configuration we estimate that thermal radiation $Q_{\textrm{rad}}$ between the post to the 50K stage dominates over thermal conductivity $Q_{\textrm{cond}}$ through the Bluestone post, with $Q_{\textrm{rad}}/Q_{\textrm{cond}}\approx5$ for measurements in which the temperature of $T_{\textrm{top}} $ is close to room temperature. We therefore performed a 3D grey\nobreakdash-body finite\nobreakdash-element simulation of the experiment to obtain $\rho_{\textrm{post}}$. Instead of using Eq.~(\ref{eq:conduction_post}), we model the heat transfer with $Q= Q_{\textrm{cond}}+Q_{\textrm{rad}}= A_{\textrm{post}}/l_{\textrm{post}} c(T)(T_{\textrm{top}}-T_{\textrm{base}})+\lambda(T_{\textrm{top}}^4-T_{\textrm{base}}^4) $ to take both heat transfer mechanisms into account. We fitted this model to our measured data assuming $c(T)$ to be constant value $c$ and use $\lambda,c$ as fit parameters. An accurate determination of the Temperature dependent thermal conductivity $c(T)$ in this temperature range would require a dedicated measurement setup optimized for a more precise characterization of $\lambda$.

Between $4 \textrm{\,K}$ and $50 \textrm{\,K}$, the temperature range in which we have not characterized the material, we assume a similar behaviour for the thermal conductivity of Bluestone as extracted from the measurements at lower temperatures. We interpolate the thermal conductivity with a polynomial of the form $\rho_{\textrm{post}}(T)=a+bT^c$ to obtain a piece\nobreakdash-wise connected function from base to room temperature [Fig.~\ref{fig:bluestone2}(c)].

We further characterize other support post materials with the same method [dots in Fig.~\ref{fig:bluestone2}(c)] and compare the values to the thermal conductivity values cited in the literature (dashed lines), observing that the deviation between literature values and measured thermal conductivity of PEEK, Nylon and Macor for temperatures up to $4\textrm{\,K}$ is less than $60\%$. The good agreement shows that the approximation of the posts as a 1D object and the analysis serve as a reliable method for measuring the thermal conductivity of Bluestone. Therefore, it is not necessary to measure every post type (50K to base) individually.

In Table~\ref{tab:bluestone_summary} we list thermal and mechanical properties of Bluestone and selected materials typically used for support structures in cryogenic systems. As mentioned in the main text, we chose to use Bluestone for the support posts for several reasons. First, Bluestone has the highest ratio of $\sigma / \rho $ below $3\textrm{\,K}$, we use this ratio as a guide for the choice of material. It shows how small the heat current from the hotter to the colder temperature stage can be under the constraint to withstand mechanical stress.
Second, it is 3D printable and enables creating geometries which are difficult to manufacture with CNC machines. Third, the material is less brittle than Macor, which means it is more robust against shocks and makes the transportation, assembly and handling of shield modules easier.

\begin{center}
\begin{table}[b]
{\footnotesize{}}
    \begin{tabular}{|l c c c c c|} 
        \hline
        Material&  SS&  M&  PEEK&  Nylon&  BS  \\
        \hline \hline
        Yield strength, $ \sigma$ MPa & 550 & 350 & 120 & 120 & 120 \\
        \hline
        Thermal cond., $\rho(1\textrm{K}) $ mW/(Km) & 70 & 5 & 5 & 3 & 1 \\ 
        \hline
        Ratio, $\sigma/\rho(1\textrm{K}) $ GNK/Wm & 8 & 70 & 20 & 40 & 120 \\
        \hline
    \end{tabular}
{\footnotesize{}\caption{\label{tab:bluestone_summary}
Properties of candidate materials for the fabrication of support posts. SS: stainless steel; M: Macor; PEEK: polyether ether ketone; N: glass- and carbon-fiber-reinforced polyamid (GF/CF Nylon); BS: Bluestone.}
}
\end{table}
\end{center}

\subsubsection{Simulations of post heat load}
The thermal model of the cryogenic link estimates the heat load of each post as a function of temperature of the enclosing stage [see Table~\ref{tab:all_parameters} in Appendix~\ref{thermal_parameters}]. For this purpose, we simulate the heat current $P_{\textrm{post},n}$ flowing through the Bluestone support posts on temperature stage $n$ using a finite element method. 
As boundary conditions we fix the base of the post to the temperature of stage $n$. The top of the post is fixed to the temperature of the stage $n+1$ and its surface is also exposed to thermal radiation from stage $n+1$. For the simulations we make use of the thermal conductivity of Bluestone $\rho_p(T)$ discussed above. In the simulation we sweep the temperature of stage $n+1$ within the typical range of temperatures observed for the cryogenic link.  

\begin{figure}[b]
	\begin{center}
		\includegraphics[width=1.0\columnwidth]{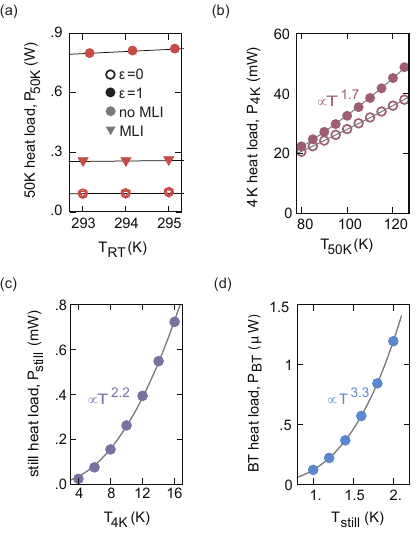}
		\caption{
			(a) Simulated heat load of a post on the 50K stage {\it vs} temperature of the vacuum can, based on finite elements method. Simulations for (b) 4K, (c) still and (d) base stage, see panel in (a) for the legend. The solid lines are fits to a power law dependence between the heat load and temperature on the top of the post. 
		}
		\label{fig:bluestone3}
	\end{center}
\end{figure}

For the 50K post we set $T_{n+1}$ to common laboratory room temperatures between $293\textrm{\,K}$ and $295\textrm{\,K}$ and simulate the heat load for one post against room temperature [see markers in Fig.~\ref{fig:bluestone3}(a)]. 
If the post is a black body absorber ($\varepsilon = 1$) and fully exposed to thermal radiation, the heat load is close to $0.8\textrm{\,W}$ per post. However, in the cryogenic link we cover the base of the post, which is half of its surface area, with high reflective MLI. After taking this into account in our simulation, the heat load decreases to $0.26\textrm{\,W}$. A fully reflective post would lower the heat load to $0.1\textrm{\,W}$.  

On the 4K stage, the absorption of black body radiation by the post has a much smaller effect and is on the order of $\approx 20\%$ of its total heat load if the 50K stage has a temperature of $120\textrm{\,K}$ [see Fig.~\ref{fig:bluestone3}(b)].  For the thermal model of the link we approximate the heat load of the post by a power law $P\propto T^{1.7}_{\textrm{50K}}$. At still and base, the radiative component can be neglected as discussed in the main text, and therefore the heat load does not depend on the emissivity of the post material. Again, both curves can be well approximated by a power law [see Fig.~\ref{fig:bluestone3}(c),(d)].

\section{THERMAL CONDUCTIVITY MEASUREMENT OF PURE COPPER TYPES}
\label{RRR_copper}

For optimizing the heat flow along the cryogenic link (see Sec.~\ref{sec:modular_design_concept} in the main text) it is critical to assert that highly conductive copper is used for the radiation shields. 

\begin{figure}[t]
	\begin{center}
		\includegraphics[width=1.0\columnwidth]{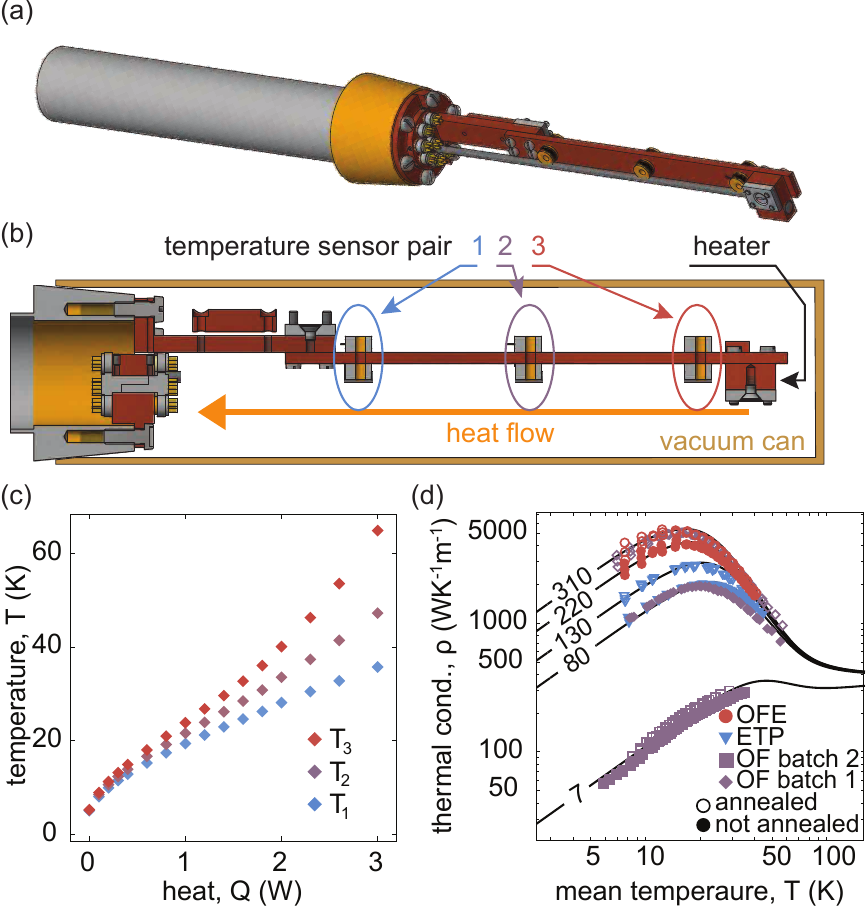}
		\caption{
			(a) Perspective view, and (b), cross-section of a CAD representation of the dipstick test setup used to characterize the thermal conductivity of copper strips. (c) Mean temperature of each temperature sensor pair at steady state {\it vs} applied power $Q$ on the heater  for a non-annealed OF batch 1 copper test sample. (d) Thermal conductivity {\it vs} temperature for different copper types before and after annealing. The black lines are theoretical curves of the thermal conductivity of copper for various RRR values.
		}
		\label{fig:RRR_measurement}
	\end{center}
\end{figure}

We have built a dipstick device with an evacuated sample space which can be immersed into liquid helium, see Figure~\ref{fig:RRR_measurement}(a,b). It allows one to assess the thermal properties of copper test strips down to 4.2~K. We attach one end of the test strip, with a size of $ 12\textrm{\,mm} \times 150\textrm{\,mm} $ and a thickness of $1,2 \textrm{ or } 3\textrm{\,mm} $, to the cold head of the dipstick. We place a heater at the other end of the test strip to induce an adjustable heat load. The head of the dipstick is enclosed in a vacuum can and pumped to a pressure of $\sim 10^{-4}\textrm{\,mbar}$ to suppress heat exchange through gas. We placed resistive temperature sensors based on ceramic oxynitride (CX1010 from LakeShore Cryotronics, Inc.) at the location labeled 1, 2, 3 on the copper test strip with equal relative distance. At each location, we have a pair of two sensors for redundancy. 

To measure the thermal conductivity, we continuously monitor the temperatures while the dipstick is immersed in liquid helium. Once all the sensors show the same temperatures, thermal equilibrium is reached and we start applying the heat $Q$ to the high temperature end of the sample. 
For each $Q$, when the temperatures are at steady state, we average the temperature of every sensor pair. Then we increase $Q$ to the next higher value. 

We observe an overall increase in measured temperature with increasing input power $Q$. The explored temperature range spans from $4.2\textrm{\,K}$ to $\sim60\textrm{\,K}$ for input powers of up to $3\textrm{\,W}$, see Fig.~\ref{fig:RRR_measurement}(c). Sensor pair 3 shows the highest temperatures as it is the closest to the heater. At larger values of $Q$, the temperature difference between the sensor pairs increase because of the finite thermal resistance of the copper strip. The radiated power of copper with the emissivity $\epsilon_{\textrm{Cu}}\simeq 0.03$~\cite{Lienhard2020} is on the order of $10\textrm{\,mW/m}^2$ for the highest temperature. The test strip has an area of $\approx0.004\textrm{m}^2$ and therefore the radiation of $\approx 40\,\mu\textrm{w}$ is negligible compared to the applied heat load of $\geq 100\textrm{\,mW}$.

As in the thermal model for the heat load calculations of the support posts (see Appendix~\ref{thermal_model}), we model the copper strip as a one dimensional bar. To extract the thermal conductivity $\rho(T)$ for a given applied heating power $Q$, we integrate Fourier's law considering two neighboring sensor pairs $i$ and $j$

\begin{equation}
    Q = \frac{A}{l_{i,j}}  \int_{T_{i}}^{T_{j}}\rho(T) dT,
    \label{eq:RRR_integral}
\end{equation} 
with the cross section area of the copper strip $A$, the distance between the temperature sensor pairs $l_{i,j}$ and the average temperature of a sensor pair $T_i$.
Expanding $\rho$ in Eq.~\eqref{eq:RRR_integral} to the first order around the mean temperature $\Bar{T}:=(T_i+T_j)/2$ allows for an approximation of the thermal conductivity of the test strip

\begin{equation}
    \rho(\Bar{T}) \simeq \frac{l_{i,j}Q}{A\Delta T_{i,j}} ,
    \label{eq:RRR_approx}
\end{equation}
using $\Delta T_{i,j}:=T_j-T_i$.

We measured more than 60 test strips of the three different copper types and the three different thicknesses of the radiation shields. The average measured thermal conductivity sorted by copper type is shown in Figure~\ref{fig:RRR_measurement}(d). The OFE copper reliably shows the highest thermal conductivity. For OF copper we see a large variations in the thermal conductivity on the material of nominally identical quality delivered to us but purchased at different times.
From the chemical composition measurements of the two OF copper batches, we hypothesize that phosphorus impurities with a mass fraction of $ \approx20\textrm{\,ppm}$ leads to the low thermal conductivity of batch 2~\cite{Kurpiers2019a}. For the other two copper types, the variation in RRR between test strips was smaller ($\leq10\%$). In the industry standards for the presented copper types not all impurities are captured, so it may be advantageous to purchase and test the raw materials before further processing for use in an experimental apparatus. A method to additionally increase the thermal conductivity of copper is to anneal the material in a vacuum oven after the machining is completed. We subsequently remeasured the thermal conductivity and observed an increase for all test strips, except for the first OF batch [see open markers in Fig.~\ref{fig:RRR_measurement}(d)].

The leading contribution to the thermal conductivity for copper originates from free electrons. At low temperatures, they predominantly scatter on impurities, whose concentration does not depend on temperature. At higher temperatures, the electron\nobreakdash–phonon scattering becomes dominant due to the increasing number of thermal phonons. As a result, the thermal conductivity reaches a maximum around $20\textrm{\,K}$~\cite{Enss2005}. A reliable empirical model to describe the thermal conductivity of copper from millikelvin to room temperature is parametrized by the residual resistivity ratio (RRR) from NIST~\cite{Simon1992}~(page 7-16). The quantity RRR is defined as the ratio of the electrical resistivity at $ 273\textrm{\,K} $ to the resistivity $ 4\textrm{\,K} $ and serves as a measure for the purity of the material.
We estimate the RRR value of a copper test strip by minimizing the quadratic error between the RRR model $\rho_{\textrm{RRR}}(\Bar{T})$ from~\cite{Simon1992} and the measured thermal conductivity $\rho(\Bar{T})$ [see black lines in Fig.~\ref{fig:RRR_measurement}(d)]. 

We observe that the OFE copper strips consistently show the highest RRR values of all selected materials. For this reason OFE copper is our material of choice for the radiation shields and thermal braids.

\section{THERMAL PARAMETERS OF THE LINK}
\label{thermal_parameters}

We characterized the thermal properties relevant for the heat transfer in the thermal model (Appendix~\ref{thermal_model}) of the link in a set of independent measurements. The key parameters are summarized in Table~\ref{tab:all_parameters}. We measured the cooling power $P_0$ of the nodes using the same methods as in~\cite{Krinner2019}. The measurements consist of applying an electrically created heat load to a given temperature stage and recording the temperature increases at steady state. The applied heat load corresponds to the cooling power on that stage. The data is then fitted with a power law. The geometric dimensions in the table are obtained from the CAD model. Parameters such as heat load of posts and thermal conductivity of copper for the radiation shields are described in Appendices~\ref{thermal_properties_bluestone},~\ref{RRR_copper}. The methods to measure thermal resistance of braid modules and radiation heat flux are described in Appendix~\ref{thermal_properties_link}. For the node and cooling unit resistance $R_\textrm{n},R_\textrm{CU}$ we used a similar method as for the braid module. The effective heat load includes the radiation heat flux, the heat load from the posts, the post density and the radiation shield geometry.  
 
\begin{center}
\begin{table*}
{\footnotesize{}}
    \begin{tabular}{|l | c c c c |} 
        \hline
         Temperature stage  &  50K&  4K&  still&  Base temperature  \\
        \hline \hline
         Reference temperature  $T_0 $ & $50\textrm{\,K}$ & $4\textrm{\,K}$& $1\textrm{\,K}$ & $10\textrm{\,mK}$  \\ 
        \hline
         Cooling power  $ P_0 $ & $36(\tilde{T}-0.75)^{0.7}\textrm{\,W} $ & $2.5(\tilde{T}-0.75)^{1.2}\textrm{\,W} $  & $80(\tilde{T}-0.5)^{2}\textrm{\,mW} $  & $4\tilde{T}^{2} \,\mu\textrm{W}$  \\
        \hline
        Post heat load  $P_{\textrm{post}} $   & $0.26 \textrm{\,W}$ & $10\tilde{T}_h^{1.7}\textrm{\,mW} $ & $34\tilde{T}_h^{2.2} \,\mu\textrm{W}$ & $10\tilde{T}_h^{3.3} \textrm{\,nW} $   \\
        \hline
        Radiation heat flux  $\varphi_{\textrm{rad}} $ & $\leq1.1 \textrm{\,W/m}^2$  & $4\tilde{T}_h^4 \textrm{\,mW/m}^2$ & $0$ & $0$ \\
        \hline
        Effective heat load  $\varphi_{\textrm{eff}} $   & $2.2 \textrm{\,W/m}^2$  & $(30\tilde{T}_h^{1.7}+4\tilde{T}_h^4) \textrm{\,mW/m}^2$ &  $150\tilde{T}_h^{2.2} \,\mu \textrm{W/m}^2$ & $1\tilde{T}_h^{3.3} \,\mu \textrm{W/m}^2$  \\
        \hline
         Effective heat load  $\psi_{\textrm{eff}} $ & $1.4 \textrm{\,W/m}$ & $(13\tilde{T}_h^{1.7}+2\tilde{T}_h^4) \textrm{\,mW/m}$ & $50\tilde{T}_h^{2.2} \,\mu \textrm{W/m}$ & $0.2\tilde{T}_h^{3.3} \,\mu \textrm{W/m}$ \\
        \hline
        Radiation shield cross\nobreakdash-section  $A$ &  $2000\textrm{\,mm}^2$ &  $1000\textrm{\,mm}^2$ &  $370\textrm{\,mm}^2$ & $200\textrm{\,mm}^2$  \\
        \hline
        Radiation shield length  $l$ &  $2250\textrm{\,mm}$ &  $2300\textrm{\,mm}$ &  $2350\textrm{\,mm}$ & $2450\textrm{\,mm}$  \\
        \hline
        Radiation shield circumference  $C_n$ &  $265$(VC)/$205\textrm{\,mm}$ &  $155\textrm{\,mm}$ &  $110\textrm{\,mm}$ & $55\textrm{\,mm}$  \\
        \hline
        RRR of the radiation shield  $\rho_{\textrm{RRR}}$ & $210$  & $230$ & $200$  & $150$  \\
        \hline
        Thermal resistance node  $R_\textrm{n}$ & $0.5/\tilde{\rho}_{75}(T) \textrm{\,K/W}$  & $2.4\tilde{T}^{-2}\textrm{\,K/W}$ & $60\tilde{T}^{-2}\textrm{\,K/W}$ & $5\tilde{T}^{-2}\textrm{\,K/mW}$   \\
        \hline
        Thermal resistance cooling unit  $R_\textrm{CU}$ & $0.2/\tilde{\rho}_{75}(T) \textrm{\,K/W}$  & $0.5\tilde{T}^{-1}\textrm{\,K/W}$ & $2\tilde{T}^{-1}\textrm{\,K/W}$ & $1\tilde{T}^{-1}\textrm{\,K/mW}$   \\
        \hline

        Thermal resistance braid module  $R_{\textrm{BM}}$& $0.15/\tilde{\rho}_{320}(T) \textrm{\,K/W}$ & $0.5\tilde{T}^{-2.6}\textrm{\,K/W}$ & $10\tilde{T}^{-1.7}\textrm{\,K/W}$  & $40\tilde{T}^{-2}\textrm{\,K/mW}$   \\
         \hline
    \end{tabular}
{\footnotesize{}\caption{\label{tab:all_parameters} Summary of extracted thermal parameters. Here, $T$ is the temperature of stage $n$, and $T_h$ is the temperature of stage $n+1$. Each quantity $\tilde{X}$ is expressed with its unit-less, renormalized quantity $\tilde{X} := X(T)/X(T_0)$ and RRR stands for residual resistivity ratio.}
}
\end{table*}
\end{center}

\section{THERMOMETRY}
\label{thermometry}

\begin{figure}[b]
	\begin{center}
		\includegraphics[width=1.0\columnwidth]{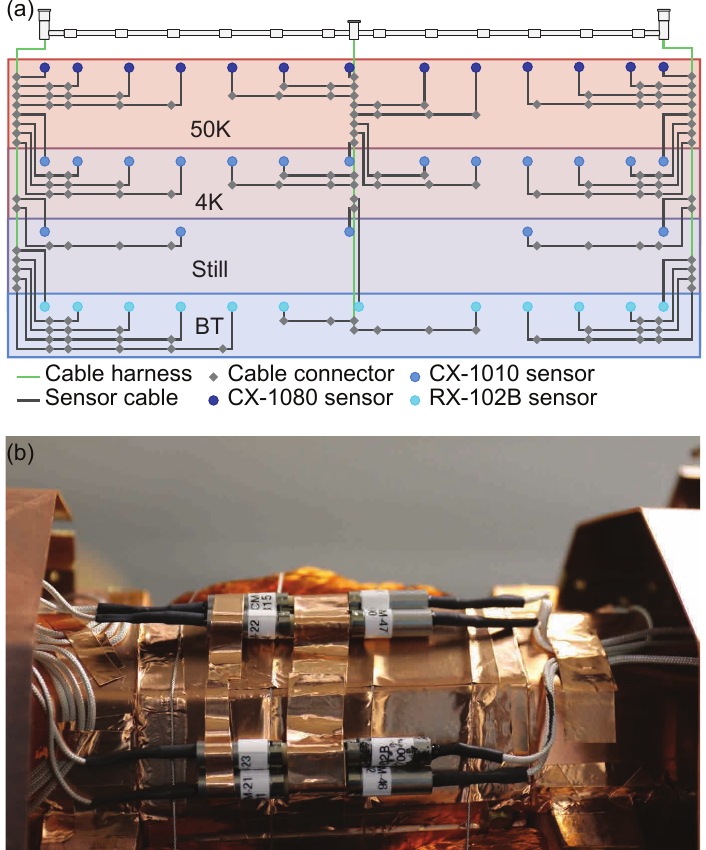}
		\caption{
			(a) Schematic illustrating the location of temperature sensors and the routing of the sensor wiring in the $30 \textrm{\,m}$ cryogenic link. (b) Photograph of a braid module at the base temperature stage with sensor wiring attached to it.
		}
		\label{fig:sensor_cabling}
	\end{center}
\end{figure}

In the following, we discuss technical aspects related to the measurement of the temperatures in the cryogenic system, shown in Fig.~\ref{fig:length_limits} in the main text.
The placement of the temperature sensors and the routing of the wiring needed to operate the sensors in the 30~meter~long cryogenic link is shown in Fig.~\ref{fig:sensor_cabling}(a). We determine the temperatures using four\nobreakdash-terminal measurements of three types of resistive temperature sensors based on ceramic oxynitride (CX-1010 and CX-1080) or ruthenium dioxide (RX-102B), each type is suited for a specific temperature range~\cite{lakeshore}. The measured resistance is converted to a temperature using an individual calibration lookup table for each sensor.

The sensor temperatures are automatically acquired using resistance bridges with a 16-channel scanner installed at each of the three cryostats. To connect to each sensor for a four\nobreakdash-terminal sensing, we use a wiring harness routed from the vacuum chamber to the different temperature stages inside the cryostat [Fig.~\ref{fig:sensor_cabling}(a)]. 

The sensor cables in the link are routed from the temperature sensor to the corresponding cable harness along and on top of the radiation shields. For ease of installation and service, each cable is divided into segments, roughly corresponding to the lengths of the radiation shields of the individual link elements. This approach allows us to test the cables installed on the individual modules before assembling the full system, and to more easily replace broken cable segments. The temperature sensor cable connectors are fixed to the radiation shields of the braid modules using copper adhesive tape [see Fig.~\ref{fig:sensor_cabling}(c)]. The cables are a few centimeters longer than the link module to compensate for the thermal contraction between the elements during a cooldown. The sensors themselves are screwed to the radiation shields.

\section{COOLDOWN DYNAMICS}
\label{cooldown_dynamics}

Cooling the entire 30\nobreakdash-meter\nobreakdash-long cryogenic link system from room temperature to $15\,$mK at the base temperature stage of the nodes takes around 6.5 days. The cooldown process of the cryogenic link can be divided into two phases, similar to the ones in a cooldown of standard dilution refrigerator.

\begin{figure}[t]
	\begin{center}
		\includegraphics[width=1.0\columnwidth]{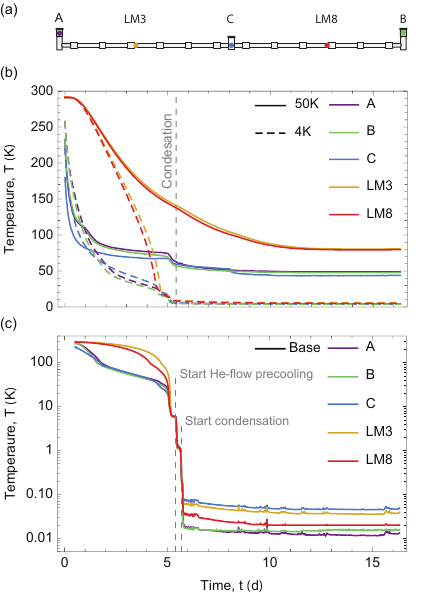}
		\caption{
			(a) A schematic of the 30-meter-long link with coloured dots indicate the positions of the temperature sensors discussed in the following panels. (b) Measured temperatures of the 50K (solid lines) and 4K (dashed lines) stages as a function of time from the beginning of a cooldown. The labels and colours correspond to the sensor positions in panel (a). (c) Base-stage temperatures during the cooldown measured at the same positions as in (b) The vertical grey lines indicate different phases during the cooldown and are discussed in the text.
		}
		\label{fig:cooldown_dynamics}
	\end{center}
\end{figure} 

In the first phase, after evacuating the vacuum container to a pressure below $10^{-4}\,$mbar, the system is cooled using pulse tube coolers at nodes A, B and C. We monitor the cooldown dynamics of the 50K and 4K stages with temperature sensors at the cooling nodes and the link modules, which are the farthest away from the heat sink of the pulse tubes (LM3, LM8). The locations of the sensors are indicated in Figure \ref{fig:cooldown_dynamics}(a). Due to the finite thermal conductivity of the radiation shields, we observe that the temperatures at the heat sinks (A-C) decrease significantly faster than the ones in the link modules, see Figure \ref{fig:cooldown_dynamics}(b).

About 4 days after starting the pulse-tube coolers, the temperature difference between the heat sinks A--C and the link modules LM3, LM8 decreases progressively faster, dropping from roughly $100\,$K to almost zero. This behaviour can be understood from the thermal diffusivity $\kappa$ of copper, defined as the ratio of thermal conductivity and specific heat, which reflects how fast it responds to a temperature change. Between $100\,$K and $10\,$K the thermal diffusivity increases by several orders of magnitude~\cite{Duthil2014}. This increase of $\kappa$ causes even small temperature gradients to be equalized quickly and leads to the observed rapid reduction of the temperature differences along the 4K stage.

The still and base temperature stages are thermally coupled to the 4K stage via heat switches during the first phase of the cooldown. They reach a quasi-steady temperature around $6\,$K after 5.5 days of cooling, as shown in Fig.~\ref{fig:cooldown_dynamics} (c) for the base stage. 

This stabilization marks the transition to the second phase of the cooldown, in which the helium heat switches at nodes A and B are deactivated and the mechanical heat switches at node C are disconnected. The helium-flow precooling process is then initiated for further cooldown. In this period, a few tens of millibar of the $^3\textrm{He}/^4\textrm{He}$ mixture are circulated through the dilution unit of each of the two nodes. Through evaporative cooling, the still and base temperature stages cool to $\sim1\,$K within one to two hours. After that, we start the condensation, where $^3\textrm{He}/^4\textrm{He}$ mixture is gradually added to the dilution unit and condenses in the mixing chamber at the base stage. After 4 hours, the mixture is fully condensed and the base temperature stage in the nodes reaches $\sim20\,$mK. Over another 12 hours the temperature lowers to a stable level of 10 to $15\,$mK. At this point, the quantum devices in the system can be operated.

In the days after the condensation, the 50K stage of the link modules still continues to cool down further, see Fig.~\ref{fig:cooldown_dynamics} (b). This subsequent and slow change is a result of the low diffusivity $\kappa$. Due to the cascaded thermal coupling to the lower temperature stages, this also affects the 4K and still stages to a small extent. After about 14 days, the whole system reaches steady state.


\end{document}